\documentclass[prd,aps,twocolumn,floatfix,nofootinbib]{revtex4-1}
\usepackage{amssymb,graphicx}
\usepackage{epsfig}
\usepackage[usenames]{color}
\usepackage{multirow}

\newcommand{\beqn}{\begin{eqnarray}}
\newcommand{\eeqn}{\end{eqnarray}}
\newcommand{\beq}{\begin{equation}}
\newcommand{\eeq}{\end{equation}}

\begin{document}

\title{Hydrodynamics in full general relativity with conservative AMR}

\author{William E.\ East${}^1$, Frans Pretorius${}^1$, and Branson C.\ Stephens${}^2$}
\affiliation{
${}^1$Department of Physics, Princeton University, Princeton, New Jersey 08544, USA. \\
${}^2$Center for Gravitation and Cosmology, University of Wisconsin-Milwaukee, Milwaukee, Wisconsin 53211, USA.}

\begin{abstract}
There is great interest in numerical relativity simulations involving matter due to the 
likelihood that binary compact objects involving neutron stars will be detected by gravitational 
wave observatories in the coming years, as well as to the possibility that binary compact object mergers could explain 
short-duration gamma-ray bursts.  We present a code designed for simulations of hydrodynamics 
coupled to the Einstein field equations targeted toward such applications.  This code has recently 
been used to study eccentric mergers of black hole-neutron star binaries.  We evolve the fluid 
conservatively using high-resolution shock-capturing methods, while the field equations are solved in 
the generalized-harmonic formulation with finite differences.  In order to resolve the various scales 
that may arise, we use adaptive mesh refinement (AMR) with grid hierarchies based on truncation error 
estimates.  A noteworthy feature of this code is the implementation of the flux correction algorithm 
of Berger and Colella to ensure that 
the conservative nature of fluid advection 
is respected across AMR boundaries. We present various tests to compare the performance of different 
limiters and flux calculation methods, as well as to demonstrate the utility of AMR flux corrections.  

\end{abstract}

\maketitle

\section{Introduction}
\label{intro}

The interface between strong field gravity and matter dynamics
promises to be one of the important frontiers in the coming years.
A new generation of gravitational wave detectors (LIGO~\cite{LIGO}, 
GEO~\cite{GEO},TAMA~\cite{TAMA}, and VIRGO~\cite{VIRGO}) are now operational, 
and within the next few years are expected to reach sensitivities that will 
allow observations of the Universe in gravitational radiation for the first 
time.  The prime targets of these observations are compact object (CO) binaries 
composed of combinations of black holes (BHs) and neutron stars (NSs).  
Modeling of such sources is a crucial ingredient to realize the promise of 
gravitational wave astronomy.  Even if an event is detected with a high 
signal-to-noise ratio, reconstructing the dynamics of the system that produced
the signal cannot be done directly
but instead will require template banks of theoretical waveforms informed 
by numerical simulations.

Compact object mergers involving NSs are expected to be significant 
sources of not only gravitational radiation, but also possible progenitors
for short-gamma-ray bursts (SGRBs)~\cite{npp92,2005Natur.437..851G,2005ApJ...630L.165L} 
and other electromagnetic and neutrino counterparts~\cite{Metzger:2011bv}.
Efforts are already underway to use potential gravitational wave sources
as triggers for searches for electromagnetic transients~\cite{LIGO_EM,LIGO_EM2}.
Observations would help constrain evolutionary models for the progenitor 
stars and their environments. Perhaps most intriguingly, the observations 
would give clues to the equation of state (EOS) of matter at nuclear densities 
(as in NS interiors), which cannot be
probed in laboratories on Earth and is not fully understood
at the theoretical level (for a broad
discussion see for example~\cite{glendenning}). The reason that the gravitational
wave signature could contain information about the matter EOS (and
other details about the internal structure of neutron stars)
is that the EOS in general has a significant effect on 
the bulk motion of matter, 
and it is this bulk motion that is the mechanism by which gravitational
waves are produced. Several studies to date have looked into this
issue, suggesting the imprint of the EOS on the gravitational
waves will be strong enough to detect~\cite{Kiuchi:2011re,Pannarale:2011pk,jocelyn,Oechslin:2007gn,Tsui:2006tr,Kokkotas:2005vr,Shibata:2005xz,Shibata:2005ss,Faber:2004fs,Bejger:2004zx,Lai:1996sv,Xing:1994ak}
(though, in some cases, the expected frequencies are higher than the range to which 
the current generation of ground-based detectors are most sensitive, thus
limiting the information which can be extracted).  
While CO binaries containing NSs are a particularly interesting class of 
sources involving general relativistic (GR) hydrodynamics, they are 
by no means the only such systems.
Examples of additional systems that have already been considered include
BH accretion tori~\cite{montero,korobkin,2011PhRvD..84b4024F,2012ApJ...744...45B} 
and NS-white dwarf mergers~\cite{vasileos}.

Thoroughly modeling systems like those described above would require evolution of the 
spacetime, the photon and neutrino radiation fields, and the magnetized, relativistic fluid.
Even a minimalistic treatment, with the Einstein equations 
coupled to the equations of relativistic hydrodynamics, represents a complex, nonlinear 
system of partial differential equations.   Numerical simulations are thus essential for 
exploring such strong field, dynamical systems. There is a long history of adapting successful
techniques for simulating Newtonian hydrodynamics to relativistic 
and general relativistic fluids which we will not attempt to summarize 
(see~\cite{grhydro_lr} for a review of general relativistic hydrodynamics).  
Instead, we will briefly attempt to place the code described in the present paper in the 
context of other recent codes developed for fluids on evolving spacetimes.
\footnote{Note that our focus is restricted to codes which handle dynamically
evolving gravitational fields.  Such codes, however, frequently owe much to earlier,
fixed-background evolution codes (see~\cite{grhydro_lr}).  In addition, advancements
such as GR-hydro with multipatch grids~\cite{thor} and with GPUs~\cite{horizon} have
recently been made with fixed-background codes.}

Several of these codes~\cite{mhdcode1,SACRA,Giacomazzo:2007ti,Jena_nsns,Bode:2009mt} solve the field equations in 
the BSSN formulation~\cite{sn,bs}.  The remainder~\cite{matt,Anderson:2006ay} use the 
generalized-harmonic formulation~\cite{garfinkle,gh3d} which we also employ; unlike 
our code, however, these groups convert to a fully first-order 
formulation~\cite{new_lindblom_et_al}.  Most groups use finite-difference methods for 
the metric evolution and a conservative, high-resolution shock-capturing (HRSC) scheme
for the hydro evolution; these unigrid algorithms are then interfaced with some sort 
of adaptive mesh refinement (AMR). A notable exception for the metric evolution is~\cite{matt}, 
which employs pseudospectral methods for the metric and then interpolates 
to a finite-volume grid for the fluid. 

Some groups have implemented the MHD equations in full GR; since these codes all make use
of conservative HRSC methods, they may be principally differentiated by how they meet the 
challenge of preserving the $\nabla \cdot {\mathbf B} = 0$ constraint.
(A straightforward finite-difference evolution of the magnetic field would generically lead to 
magnetic monopoles and, hence, unphysical behavior.)  
{\tt WhiskyMHD} employs constrained transport~\cite{Giacomazzo:2007ti} for this purpose, which 
preserves the constraint to machine accuracy, whereas the code of~\cite{chawla} 
uses hyperbolic divergence cleaning.  Constrained transport, however, requires special
interpolation at refinement level boundaries in order to preserve the constraint. 
The Illinois group found that a vector-potential 
formulation of the MHD equations works well when coupled to AMR~\cite{illinoisNewMHD}.  
This is because the constraint is preserved by construction with the 
vector-potential, even with the restriction and prolongation operations of AMR (see 
also~\cite{illinoisMHD2} for a thorough examination of the electromagnetic 
gauge condition).  Studies indicate that magnetic fields do not
significantly affect the gravitational dynamics of CO mergers (see {\em e.g.} \cite{chawla}), but they 
could be critical for understanding EM counterparts including the possible formation of a SGRB 
engine. A new method to treat the MHD equations was recently presented in~\cite{lehner_new}, where ideal MHD is used in high matter density regions ({\em e.g.} inside a NS), while the force-free 
approximation is used elsewhere ({\em e.g.} the magnetosphere of a NS). The authors applied the method to study the
collapse of magnetized hypermassive NSs (which could be formed via binary NS mergers) and suggested that 
intense EM outbursts could accompany such events.

Besides MHD, the other major advances in the physical model for numerical relativity codes 
have been in the arena of microphysics.  
While the $\Gamma=2$ EOS was the community standard for quite some time, most codes now allow for
a nuclear theory-based EOS~\cite{matt2,shibataBNS} and/or use various parametrized, piecewise polytropic 
EOSs inspired by the range of plausible nuclear EOSs~\cite{shibataBHNS4,illinoisWDNS}.  These advances in EOS 
description primarily affect the cold NS structure, but the group developing the {\tt SACRA} code has 
also begun to account for neutrino transport via a simplified leakage 
scheme~\cite{Kiuchi:2011re,2011PhRvL.107e1102S}.  The same group has also made available a
formulation for a more accurate truncated moment scheme with a variable Eddington factor 
closure~\cite{truncated_moment}, which shows much promise for numerical relativity simulations 
with neutrino physics beyond the leakage approximation.

Another category of GR hydrodynamics codes employs the
conformal-flatness approximation, which is particularly useful when
supernova simulations are the target application.  An example is 
CoCoNuT/VERTEX, which incorporates relativistic hydrodynamics, conformally
flat gravity, and ray-by-ray neutrino transport~\cite{coconut}. 
The code of~\cite{cerda_duran} employs a similar scheme for hydrodynamics
and gravity but adds a test magnetic field; this code has been used to 
study the magnetorotational instability in supernovae.

Newtonian (and semi-Newtonian)~\cite{lee_kluzniak,magma}, conformally flat~\cite{faber,oechslin},
and fixed-background~\cite{laguna} SPH codes represent an important, orthogonal approach 
to studying CO interactions.
SPH has an advantage over Eulerian schemes when a large range of spatial scales
is involved.  Such a situation may arise in CO mergers when material is stripped
from a star in a tidal interaction and forms an extended tail.  On the other hand,
Eulerian codes are the standard approach when strong shocks are present, as
would arise in binary NS mergers or disk circularization.  (Recent progress
has been made, however, in applying SPH to situations with relativistic 
shocks~\cite{Rosswog_SRHD}.)  In addition, SPH has not (to our knowledge) yet 
been coupled to a code which evolves the full Einstein equations.  Nonetheless,
comparisons between Eulerian and SPH results could prove very useful on a
problem-by-problem basis to characterize the errors in both methods.

Though current efforts in GR simulations involving matter tend to focus on 
increasingly complex physical models, there remain many unanswered questions
in the astrophysics of compact objects that can be addressed with
a code which solves the Einstein equations coupled to perfect 
fluid hydrodynamics.  We have thus focused our 
code development on hydrodynamics in full GR, while maintaining
a flexible infrastructure to accommodate additional physics modules 
in the future.  We evolve the field equations in the generalized-harmonic formulation
using finite differences.  The fluid is evolved conservatively using one of several
different shock-capturing techniques we test here. We have also implemented the hydrodynamical
equations in a manner that is independent of EOS.  We make use of AMR by dynamically adapting 
the mesh refinement hierarchy based on truncation error estimates of a select
number of the evolved variables.  We also utilize Berger and Colella~\cite{bc89}
style flux corrections (also known as ``refluxing'') in order to make the use of AMR compatible 
with the conservative nature of the hydrodynamic equations.
Though AMR flux corrections have been implemented in other 
astrophysical hydrodynamics codes (such as Athena~\cite{ATHENA}, CASTRO~\cite{CASTRO}, 
{\tt Enzo}~\cite{EnzoMHD}, and FLASH~\cite{FLASH}),
to our knowledge this algorithm has not been used previously for  
hydrodynamics simulations in full general relativity.\footnote{Note that ``flux correction'' 
here refers to the enforcement of conservation at AMR boundaries,
not the recalculation of fluxes with a more dissipative scheme to preserve stability as in 
Athena~\cite{athena_rmhd}.} 
A further noteworthy feature of our implementation is that we store corrections
to the corresponding fluid quantity integrated in the volume of a given cell 
instead of the flux, allowing for easy implementation within
a computational infrastructure that supports cell-centered but not face-centered
distributed data structures.
The code described here has recently been applied to studying BH-NS mergers with eccentricity
as may arise in dense stellar systems such as galactic nuclear clusters and globular 
clusters~\cite{ebhns_letter,ebhns_paper}.

In the remainder of this paper we outline our computational methodology for simulating hydrodynamics coupled to the Einstein field equations
and describe tests of this methodology.
In Sec.~\ref{methods} we review the generalized-harmonic approach to solving the field equations 
and present our methods for conservatively evolving a perfect fluid coupled to gravity, including 
our method for inverting the conserved quantities to obtain the primitive fluid variables and the 
implementation of flux corrections to enforce the conservation
of fluid quantities across AMR boundaries.  In Sec.~\ref{tests} we present
simulation results which test these methods, highlight the strengths and weaknesses of various shock -capturing techniques, and demonstrate the utility of the flux correction algorithm.

\section{Computational methodology}
In this section we begin by explaining the basic equations and variables we use to numerically evolve the 
Einstein equations in Sec.~\ref{efe_sec} and then discuss the conservative formulation of the hydrodynamics
equations that we use in Sec.~\ref{cons_sec}.  The evolution of conserved fluid variables necessitates an algorithm
for inverting these quantities to obtain the primitive fluid variables which we present in Sec.~\ref{inversion}. 
Finally in Sec.~\ref{fc_explain} we present the details of our algorithm for AMR with flux corrections.
\label{methods}

\subsection{Solution of the Einstein equations}
\label{efe_sec}
We solve the field equations in the generalized-harmonic formulation~\cite{garfinkle,gh3d}  where we fix the coordinate degrees of freedom by specifying the evolution of the source functions $H^a:= \Box x^a$.  In this form the evolution equation for the metric, $g_{ab}$, becomes manifestly hyperbolic:
\begin{eqnarray}
g^{cd}\partial_{c}\partial_{d}g_{ab} + \partial_{b}g^{cd}\partial_{c}g_{ad} + \partial_{a}g^{cd}\partial_{c}g_{bd} \nonumber\\
+ 2H_{(a,b)} - 2H_{d}\Gamma^d_{ab} + 2\Gamma^c_{db}\Gamma^d_{ca}\nonumber\\
= -8\pi (2T_{ab} - g_{ab}T)
\label{efe}
\end{eqnarray}
where $\Gamma^a_{bc}$ is the Christoffel symbol,
$T_{ab}$ is the stress-energy tensor, and $T$ is its trace. 
We evolve the metric, the source functions, and their respective time derivatives
using fourth-order Runge-Kutta where the spatial derivatives are calculated using fourth-order
accurate finite-difference techniques. In other words, we have reduced the evolution equations to first order
in time so that there are 28 ``fundamental'' variables $\{g_{ab}, H_a, \partial_t g_{ab}, \partial_t H_a\}$,
but we directly discretize all first and second spatial gradients without the introduction
of additional auxiliary variables.

Analytically one can show~\cite{Friedrich} that if one begins with initial data that satisfies the 
Hamiltonian and 
momentum constraints, 
initially set $H^a= \Box x^a$, and then evolve the metric
according to (\ref{efe}) and the source functions according to some specified 
differential equations, then the constraint equation $H^a-\Box x^a=0$ will be
satisfied for all time. Numerically this statement will only be true to within
truncation error, which can grow exponentially in black hole space times; to 
prevent this we add constraint damping terms as in~\cite{Gundlach:2005eh,paper2}.
In practice, ensuring that
$H^a-\Box x^a$ is converging to zero for a given numerical simulation run at different resolutions provides an excellent check that the numerical solution is indeed converging to a solution of the field equations.  

As described in~\cite{gh3d}, the computational grid we use is compactified so as to include spatial infinity.  This way
we can impose boundary conditions on the metric simply by requiring that it be Minkowski.
However we evolve the metric of the uncompactified coordinates since the compactified metric
is singular at spatial infinity.

\subsection{Conservative Hydrodynamics}
\label{cons_sec}
Coupled to gravity we consider a perfect fluid with stress-energy
tensor
\beq
T^{ab} = \rho h u^{a} u^{b} + g^{ab}P \ ,
\eeq
where $h := 1 + P/\rho + \epsilon$ is the specific enthalpy and $u^a$ is the four-velocity
of the fluid element. The intrinsic fluid quantities $\rho$, the rest-mass density; $P$, the pressure; and $\epsilon$, the specific energy are defined in the comoving 
frame of the fluid element.  The equations of hydrodynamics are then written in 
conservative form as follows~\cite{PhysRevD.61.044011}:
\beqn
\partial_t D + \partial_i (D v^i)  & = & 0  \label{mass_cons} \\
\partial_t S_{a} + \partial_i\left(\sqrt{-g}T^i{}_{a}\right) & = &  \frac{1}{2}\sqrt{-g}T^{bc}\partial_ag_{bc}  
\label{dtSmu}
\eeqn
where $v^i$ is the coordinate velocity, $g$ is the determinant of the metric,
and the index $i$ runs over spatial coordinates only.
Note that (\ref{dtSmu}) explicitly contains the time derivative of the metric
for index $a=t$.  The conserved variables $D$ and $S_{a}$  are defined as follows:
\beqn
D       & := & \sqrt{-g}\rho u^t \\
S_{a} & := & \sqrt{-g}T^t{}_{a} \label{S_a_def}
\eeqn
where $D$ is simply the time component of the matter 
4-current\footnote{In some implementations of the GR (magneto)hydrodynamic
equations, see for e.g.~\cite{2003ApJ...589..444G}, the analog of $S_t$ 
in (\ref{S_a_def}) that is evolved has the rest-mass density $D$ subtracted off.
This could provide improved results in situations where the rest-mass density
is orders of magnitude larger than the internal or magnetic energy,
and accuracy in these latter quantities is important. Though we have not explored this alternative,
in the scenarios studied here (in particular since we are not looking at the
behavior of magnetic fields) the added effect of a small amount of internal relative to rest energy 
on the dynamics of the fluid or metric will be negligible, and we expect
either definition of $S_t$ to give comparable accuracy results here.}.

In some situations
we wish to perform axisymmetric simulations where we use the symmetry to reduce
the computational domain to two dimensions.  We do this using a modification of 
the Cartoon method~\cite{Alcubierre:1999ab} as described in~\cite{gh3d}, where we take the $x$-axis as the axis of symmetry, 
and only evolve the $z=0$ slice of the spacetime.
For the hydrodynamics this means that
effectively each fluid cell becomes a cylindrical shell, and we use the fact that the
Lie derivative of the fluid fields with respect to the axisymmetric killing vector are
zero to rewrite the coordinate divergences in the above equations as
\begin{equation}
\partial_i(D v^i) = \partial_x(D v^x) + 2\partial_{y^2}(yD v^y)
\end{equation}
and similarly for $\partial_i(\sqrt{-g}T^i{}_{a})$ for the $t$ and $x$
components.
For the $y$ component there is an additional source term
\begin{eqnarray}
\partial_i(\sqrt{-g}T^i{}_{y}) = \partial_x(\sqrt{-g}T^x{}_{y}) + 2\partial_{y^2}(y\sqrt{-g}T^{y}_{y}) -\nonumber\\
(S_{z}v^z+\sqrt{-g}P/y) .
\end{eqnarray}
By writing the $y$ flux contribution in terms of $\partial_{y^2}$ we ensure that when we discretize
our evolution will be conservative with respect to the cylindrical shell volume element.
We choose a special form for the equation for $S_z$:
\begin{equation}
\partial_t S_{z} + \partial_x(\sqrt{-g}T^x{}_{z}) + \frac{2}{y}\partial_{y^2}(y^2\sqrt{-g}T^{y}_{z})=0,
\end{equation}
since in axisymmetry the quantity $yS_z$ is exactly conserved (that is, it has no source term). 

The conservative evolution system is solved numerically using HRSC schemes.   
We briefly summarize the different methods we have implemented and test in this paper, 
though the references should be consulted for more complete details.
For calculating intercell fluxes we have implemented
HLL~\cite{hll}, the Roe solver~\cite{eulderink}, and the Marquina flux~\cite{marquinaFlux} method. 
The HLL method is straightforward to implement since it does not require the spectral decomposition of the
flux Jacobian and is based on estimates for the largest and smallest signal velocities.
The Roe solver works by solving the linearized Riemann problem obtained using the flux Jacobian at
each cell interface (using the so-called Roe average of the left and right states).  The Marquina flux method 
is an extension of this idea that avoids the artificial intermediate state and switches to a more viscous local Lax-Friedrich-type method from~\cite{Shu198932} when the characteristic speeds change sign across the interface.
Since the latter two methods require the spectral decomposition of the flux Jacobian, we
give it for our particular choice of conserved variables in the Appendix. 
For reconstructing fluid primitive variables at cell faces we have implemented
MC and minmod~\cite{toro}, PPM~\cite{ppm} \footnote{In particular we use the reconstruction parameters presented in~\cite{MartiMuller}.}, and WENO5~\cite{Jiang1996202} \footnote{Specifically, we perform reconstruction with the stencils and weights presented in Section A2 of~\cite{weno5}.}, all of which may 
be used interchangeably with any flux method.  
MC and minmod are both slope limiter methods that reduce to linear reconstruction for smooth flows.  Minmod is
the more diffusive of the two.  In comparison, PPM and WENO5 are higher-order reconstruction methods.  PPM is based on 
parabolic reconstruction with modifications to handle contact discontinuities, avoid spurious oscillations from shocks by
reducing order, and impose monotonicity.  WENO5 combines three different three-point
stencils with weights that are determined by a measure of the smoothness of the quantity being reconstructed.
The specific fluid quantities that we reconstruct on the cell faces are $\rho$, $u$, and $WU^i$, 
where $u := \rho\epsilon$, $W$ is the Lorentz factor between the local fluid element and an observer normal to the 
constant $t$ hypersurfaces, and $U^i$ is the Eulerian velocity (the explicit
form of which is given in the following section).
We choose to reconstruct
$WU^i$ instead of simply $U^i$ since any finite value of this quantity corresponds to a subluminal velocity.

The fluid is evolved in time using second-order Runge-Kutta. 
Since the fluid is evolved in tandem with the metric, the first and second substeps of the fluid Runge-Kutta time step
are chosen to coincide with the first and third substeps of the metric time step.
Since the spatial discretization of the fluid equations that we use is only second-order we choose to use second-order time stepping 
for the hydrodynamics and we have not yet experimented with higher-order methods.
We still use fourth-order Runge-Kutta for nonvacuum metric evolution (even though for evolutions with matter the overall 
convergence rate will be no greater than second-order) both for convenience and because in vacuum dominated regions we may expect
some improvement in accuracy. 
For general relativistic hydrodynamics we evolve the fluid on 
a finite subset (though the majority) of the total grid (which as mentioned extends to spatial
infinity through our use of compactified coordinates), and at the outer boundary for the fluid 
we impose an outflow condition. 
 
Finally, as is common practice for this method of simulating hydrodynamics,
we require that the fluid density never drop below a certain threshold, adding a so-called 
numerical atmosphere.  We give this numerical atmosphere a spatial dependence that makes it 
less dense approaching the boundaries\footnote{Specifically we let 
$\rho_{\rm atm}(x_c,y_c,z_v)=\bar{\rho}\cos^2(x_c)\cos^2(y_c)\cos^2(z_c)$
where $\bar{\rho}$ is a constant, and $(x_c,y_c,z_c)$ are the compactified coordinates which range from -1 to 1.} 
and choose
a maximum value that makes it dynamically negligible (typically at least 10 orders of magnitude below the maximum density). 
The atmosphere is initialized using a cold equation of state ({\em e.g.} a polytropic equation of state).

\subsection{Primitive Inversion}
\label{inversion}
The set of hydrodynamical equations is closed by an EOS of the form 
$P=P(\rho,\epsilon)$.  While the conserved variables
$S_{a}$ and $D$ are simply expressed in terms of fluid primitive variables
($\rho$, $P$, $\epsilon$, and $v^i$) and the metric, the reverse is not true.  
This necessitates a numerical inversion to obtain the primitive variables 
following each update of the conserved variables. The method we use is similar to the one used in~\cite{Noble}
for spherical symmetry. First, we decompose the 4-dimensional metric
into the usual ADM space plus time form
\beqn
ds^2&=&g_{ab} dx^a dx^b \nonumber \\
    &=& -\alpha^2 dt^2 + \gamma_{ij} (dx^i + \beta^i dt) (dx^j + \beta^j dt) \label{ADM}
\eeqn
where $\gamma_{ij}$ is the spatial metric, $\alpha$ the lapse function
and $\beta^i$ the shift vector. Then, from the metric and
conserved variables we construct two quantities,
\beqn
S^2 & := & \gamma^{ij}S_iS_j = \gamma H^2 W^2(W^2-1) \label{ssquared} \\
E   & := & \beta^i S_i - S_t  = \sqrt{-g} (H W^2 - P) \ , \label{EnergyEq}
\eeqn
where $H := \rho h$ and $\gamma$ is the determinant of the spatial metric.
We reduce the problem of calculating the primitive fluid variables from
the metric and conserved variables to a one-dimensional root problem, where
we begin with a guess for $H$ and iteratively converge to the correct value
such that $f(H)=0$ for some function.  From~(\ref{EnergyEq}) we can choose 
\beq
f(H)= E/\sqrt{-g} - HW^2+P.  
\eeq
Note that given the metric and conserved variables,
$f(H)$ is only a function of $H$, and can be computed as follows.
First, calculate $W^2 = (1+\sqrt{1+4\Lambda})/2$ where
\beq
\Lambda := \frac{S^2}{\gamma H^2} = W^2 (W^2 -1) \ . \label{LambdaEq}
\eeq
Then compute $\rho$ and $\epsilon$ from
\beq
\rho=D/(\sqrt{\gamma}W),
\eeq 
and 
\beq
\epsilon = -H(W^2-1)/\rho + WE/(D\alpha) -1 \ , \label{eps_inv}
\eeq 
respectively. Once $\rho$ and $\epsilon$ are known, $P$ can
be obtained from the equation of state, and then $f(H)$ above.
An iterative procedure for solving $f(H)=0$, where $f(H)$ is
calculated as just described, thus gives the primitive variables $\rho$, $P$, and
$\epsilon$.
The three-velocity can then be computed from
\beq
U^i = \frac{\gamma^{ij}S_j}{\sqrt{\gamma}HW^2} \ , 
\eeq
where the Eulerian velocity $U^i$ is related to the grid three-velocity through
$U^i=(v^i + \beta^i)/\alpha$.  This inversion scheme is implemented so as to 
allow any EOS of the form $P=P(\rho,\epsilon)$; thus, $\Gamma$-law, piecewise 
polytrope, and tabular
equations of state such as the finite-temperature EOS of 
Shen et al.~\cite{shen1,shen2} (for a given electron fraction $Y_e$) are all supported.  

In practice 
we solve for $f(H)=0$ numerically using Brent's method~\cite{brent}, which does not require knowledge of 
derivatives and is guaranteed to converge for any continuous equation of state as long as 
one begins with a bracket\footnote{
The initial bracket for the
root finding is chosen by first checking if $[H_0/(1+\delta),H_0(1+\delta)]$,
where $H_0$ is the value of $H$ computed for the primitive variables at the previous
time step and $\delta>0$ is a parameter we take to be 0.4,
is a valid bracket around the zero of $f(H)$.  If it is not, as a failsafe we try
successively larger brackets with $[H_0/(1+\delta)^n,H_0(1+\delta)^n]$ for $n\geq2$.
}
around the correct solution. This can be useful when dealing with 
equations of state interpolated from tabulated values. 
One can avoid losing accuracy in the ultrarelativistic and nonrelativistic limit
by Taylor expanding the above inversion formulas (see~\cite{Noble}), for example,
in $1/\Lambda$ and $\Lambda$, respectively.  We have implemented such expansions
in our primitive inversion algorithm, 
though we have not yet made any significant study of the inversion calculation in these regimes. 

In some cases the conserved variables will, due to numerical inaccuracies, evolve to a state that does not correspond 
to any physical values for the primitive variables.  This causes the inversion procedure to fail.  
This can happen in very low density regions that are not dynamically important but still must be addressed.  
We handle such situations using a method similar to that of~\cite{PhysRevD.61.044011} by ignoring the value 
of $S_t$ and instead requiring the fluid to satisfy a cold equation of state.

\subsection{AMR with flux corrections}

\label{fc_explain}
Many of the problems we are interested in applying this code to involve
a range of length scales, and in many cases we expect the small length
scale features {\em not} to be volume filling, for example the individual compact
objects in binary mergers.
Such scenarios can be efficiently resolved with Berger and
Oliger style adaptive mesh refinement~\cite{bo84}. A description
of the variant of the algorithm we use can be found in~\cite{2006JCoPh.218..246P}; here we 
mention some particulars to this implementation, and give a detailed description
of the extension to ensure conservation across refinement boundaries.

The computational domain is decomposed into a hierarchy of uniform meshes, where
finer (child) meshes are entirely contained within coarser (parent) meshes. The hierarchy is
constructed using (primarily) truncation error (TE) estimates, which are
computed within the Berger and
Oliger time subcycling procedure by comparing the solution obtained on 
adjacent levels of refinement
before the coarser levels are overwritten with the solution from the finer level.
Typically we only use the TE of the metric variables, since fluid variables in general 
develop discontinuities as well as turbulent features that do not follow strict convergence.  
The layout of the AMR hierarchy is then periodically adjusted in order to keep the TE
below some global threshold.  In some situations we also require that a region where the fluid 
density is above a certain threshold always be covered by a minimum amount of resolution.  This can
be used to ensure, for example, that the resolution around a NS does not temporarily drop 
below some level even if the TE of the metric variables in the neighborhood of the star
becomes small. 

When setting the values of the metric variables on the AMR boundary of a given child level we interpolate
from the parent level using third-order interpolation in time and fourth-order in space.
For the cell-centered variables, the outer two cells in each spatial direction (for a refinement ratio of 2) 
on a child level are initially set using second order interpolation in time and space from the parent level.  
Following the evolution of the child level and flux correction applied
to the parent level when they are in sync as described below, but 
before the cell-centered values on the child level are injected into the parent level,
the values in the child boundary cells are reset using first-order 
conservative (spatial) interpolation from the parent level (\emph{i.e.} the value in the child cell is set to be the same as 
that of the parent cell in which the child cell is contained). This ensures that the boundary
cells on the child level are consistent with the corresponding flux-corrected cells on the parent level
but does not affect the order of convergence of the scheme since these values are not used in the evolution step.
During a regrid when adding cells to the domain of a refined level we also use 
first-order conservative interpolation from the overlapping parent level
to initialize the values of the fluid variables at new cells (fourth-order
interpolation is used for the metric variables).
Note that the actual domain that is refined is larger than the volume
where the TE estimate is above threshold by a given buffer in any
direction. The buffer size and regridding interval are chosen so that if change in the region
of high TE is associated with bulk motion of the solution (e.g. the NS
moving through the domain), this region will never move
by more that the size of the buffer between regrids. This ensures that new cells
(for this kind of flow) are always interpolated from regions of the parent
that are {\em below} the maximum TE threshold. Thus, though the interpolation
operation to initialize new cells is first-order, we find the error it
introduces is negligible ({\em i.e.}, below the maximum desired TE).

AMR boundaries require special treatment in conservative hydrodynamics codes however, since the fluxes
across the boundary of a fine-grid region will not exactly match the 
corresponding flux calculated on the coarse-grid due to differing truncation errors. 
To enforce conservation, we correct the adjacent coarse grid cells using the fine-grid 
fluxes according to the method of Berger and Colella~\cite{bc89}.  In the remainder
of this section we review the algorithm and outline our specific implementation.

\begin{figure*}
\begin{center}
\setlength\fboxsep{0pt}
\setlength\fboxrule{0.5pt}
\fbox{\includegraphics[width=4.0in,clip=true,viewport=0 40 900 739]{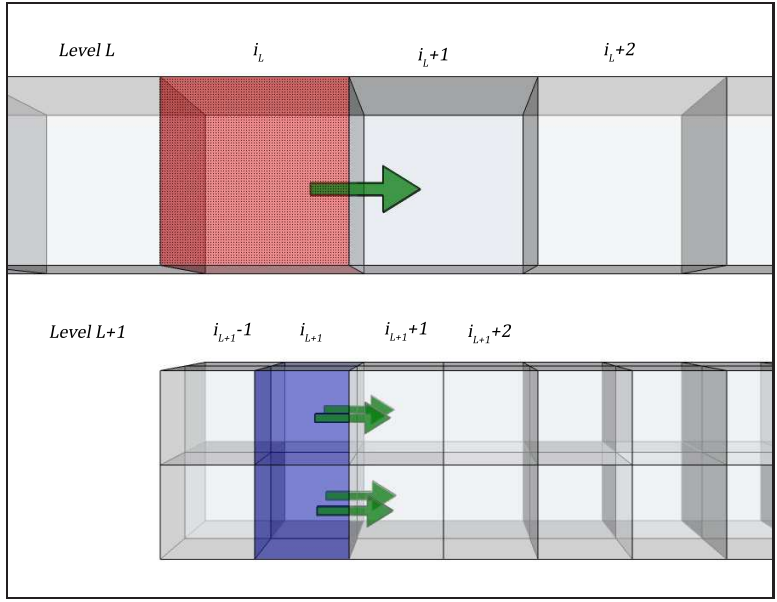}}
\caption{
A visualization of a refinement level boundary and its treatment in  the flux correction algorithm.  
The top shows cells in the $x$ direction on refinement level $L$ while
the bottom shows equivalent cells for the $L+1$ refinement level (here the refinement ratio is 2). Fluxes are
symbolized by arrows.
On the bottom level the blue cells (``type B'' in the discussion in the text) and those to the left on level $L+1$ are boundary
cells and will have their values set by interpolation from level $L$ following an evolution step on level $L$.
Because of truncation error, subsequent evolution on level $L+1$ will give a flux differing from that computed on the 
parent level $L$. Consequently, when the new fine grid solution is injected
back to the parent level (in cells to the right of the red/dotted pattern cell), the solution about the boundary will
no longer be consistent with the flux previously computed there. The correct this, the fluid quantity in the
red/dotted pattern cell is adjusted to exactly compensate for the difference in flux computed between the coarse
and fine levels.
\label{fc_diagram}
}
\end{center}
\end{figure*}

We will concentrate on the evolution of $D$ on a 3-dimensional spatial grid, though the remaining conserved fluid quantities 
are treated the same way, and modification to different numbers of spatial dimensions is trivial.   
Equation~(\ref{mass_cons}) is evolved numerically at a given resolution as
\begin{eqnarray}
 D^{n+1}_{i,j,k} = D^n_{i,j,k} &-& \nonumber \\
 \Delta t [(F^x_{i+1/2,j,k}&-&F^x_{i-1/2,j,k})/\Delta x \nonumber\\
+(F^y_{i,j+1/2,k}&-&F^y_{i,j-1/2,k})/\Delta y \nonumber\\
+(F^z_{i,j,k+1/2}&-&F^z_{i,j,k-1/2})/\Delta z]
\end{eqnarray}
where $D^{n}_{i,j,k}$ is the volume average of $D$ over the $(i,j,k)$ cell at time $t=n\Delta t$;
$F^x_{i+1/2,j,k}$ is the flux $F^x=Dv^x$ through the $(i+1/2,j,k)$ cell face; $\Delta x$ is the $x$ length of 
each cell and so on for the $y$ and $z$ direction.  In practice the flux values will be calculated
with some HRSC technique combined with Runge-Kutta, but the specifics are not relevant here.  Now consider
a situation with two sequential levels of refinement, $L$ and $L+1$, where level $L+1$ has
a higher resolution with spatial refinement ratio
of $r$ in each direction, and its domain is a subset of level $L$. (In practice, we always take $r=2$.)
Here we focus the discussion on a left boundary in the $x$ direction, as illustrated
in Fig.~\ref{fc_diagram}; boundaries along the right face and other coordinate
directions are treated in a like manner.

When evolving according to the Berger-Oliger algorithm, after each time step of length $\Delta t$ is taken on level
$L$, $r$ steps of length $\Delta t/r$ are taken on level $L+1$.  Then the results obtained on $L+1$ are
injected into level $L$ where the levels overlap {\em i.e.}, the restriction operation is performed
conservatively by setting the value in the parent cell to the (coordinate) volume-weighted average of the child cells that  
make up the parent cell.
Now on level $L$, the change in $D$ due to flux going through the cell 
face $(i_L+1/2,j_L,k_L)$ on a timestep will be 
\begin{equation}
\delta D_L = -\frac{\Delta t}{\Delta x} F^x_{i_L+1/2,j_L,k_L}(t_n).
\label{fcL_eqn}
\end{equation}
On level $L+1$, 
the change in $D$ in one fine-level time step due to flux passing through one of the $r^2$ cell faces 
that make up this same interface is
\begin{eqnarray}
\delta D_{L+1,j,k,m} =\nonumber -\frac{\Delta t /r}{\Delta x/r}\times \nonumber\\
F^x_{i_{L+1}+1/2,j_{L+1}+j,k_{L+1}+k}(t_n+m\Delta t/r).
\label{fcL1_eqn}
\end{eqnarray}
for $j$, $k$, and $m\in \{0,1,\dots,r-1\}$.
Now because of truncation error, in general the change
in the net ``mass''
\footnote{
For the conserved fluid variable $D$ which we focus on for specificity, 
the value of the quantity integrated within the volume of a cell in fact represents
the rest mass in that cell.  Throughout this section we will therefore use the term `mass' to refer to the value of a
conserved fluid variable volume integrated within a cell, though for other conserved fluid variables this will not
correspond to a physical mass.
}
$\delta M_L := \delta D_L V_L$ within the coarse-level cell
at $(i_L,j_L,k_L)$ computed with the coarse-level fluxes
will not equal the corresponding quantity
$\delta M_{L+1} := \sum_{j,k,m}\delta D_{L+1,j,k,m} V_{L+1,j,k,m}$
computed with the fine-level fluxes, 
where $V_L$ is the coordinate volume of the cell $(i_L,j_L,k_L)$ and  
$V_{L+1,j,k,m}$ is the coordinate volume of the cell $(i_{L+1},j_{L+1}+j,k_{L+1}+k)$.
Thus, after the values of $D$ on level $L+1$ are injected into level $L$ 
(in cells $(i_L+1,j_L,k_L)$ and to the right in this example), the
solution on level $L$ will suffer a violation of mass
conservation proportional to $\delta M_L - \delta M_{L+1}$.
To restore the conservative
nature of the algorithm, the idea, described in detail below, is to adjust
the conservative variable $D$ in the cell $(i_L,j_L,k_L)$ post-injection
by an amount to exactly compensate for this truncation error induced
difference.

The scheme originally proposed in 
\cite{bc89} is to define an array that keeps track of a correction to the fluxes through cell faces
on level $L$ that make up the boundary of the evolved cells on level $L+1$.  
Consider the case where $(i_L+1/2,j_L,k_L)$ is such a face.
This face-centered flux correction array, $\delta F$, 
is initialized with the inverse of the flux in~(\ref{fcL_eqn}), $\delta F=-F^x_{i_L+1/2,j_L+j,k_L+k}$, and 
then during the course of taking the $r$ time steps on level $L+1$
receives corrections from the terms in~(\ref{fcL1_eqn}) 
\beq
\delta F \to \delta F  + \frac{1}{r^3}\sum_{j,k,m} F^x_{i_{L+1}+1/2,j_{L+1}+j,k_{L+1}+k}(t_n+m\Delta t/r).
\eeq
After the cell values on level $L$ are overwritten by the injected values on level $L+1$ where 
they overlap, the cells on level $L$ that abut level $L+1$ though are not themselves covered by level $L+1$
cells are corrected with the flux stored in $\delta F$.

The way we implement the flux correction algorithm is slightly different from this.  In particular we
wish to avoid the added computational complexity of implementing face-centered grid functions, 
and therefore we keep track of a cell-centered correction. The correction is thus also more naturally represented
as a correction to the fluid quantity integrated within the volume of the cell ({\em e.g.} for $D$ the rest-mass) rather than a flux.
Again referring to Fig.~\ref{fc_diagram}, we define the first few cells at the boundary of level $L+1$ as buffer cells 
since the calculation of flux requires knowledge of the state on both sides of the interface.  
These cells will have their values set by interpolation from those in level $L$.
The innermost buffer cells for the boundary on level $L+1$ we call type B cells (blue
cells in the lower half of the figure). These are the cells where the level $L+1$ contribution to the 
mass correction will be stored.
The cell on level $L$ which contains the type B cell we will refer to as a type A cell (red, dotted-pattern cell).  Type A cells
are the ones that receive mass corrections in this algorithm.
For each cell on each refinement level we use a bitmask grid function that
indicates whether the cell is one of the above types (A or B), and if so which of
the six possible faces ($+x$, $-x$, $+y$, $-y$, $+z$, $-z$) abut the boundary.
For simplicity in the implementation we do not allow grid hierarchies where a cell would be both type A and 
type B\footnote{In other words, an inner (non-physical) boundary on level $L$ must be at least
one cell away from any inner boundary on level $L-1$. If the hierarchy is generated
by truncation error which is sufficiently smooth, inner boundaries will
typically not be coincident. Also, experience suggests it is often
more challenging to get an AMR evolution stable if inner boundaries are too close, 
so in all this restriction is not particularly limiting.}.

In the following we outline the additional tasks
relative to the basic Berger-Oliger algorithm that need to be performed with
our implementation of Berger-Colella. Following the spirit of these algorithms,
we break down the tasks into those the AMR ``driver" code implements, which do not require knowledge 
of the specific equations being evolved or what physical quantities the variables represent,
and conversely the ``application" steps that would need to be implemented by a
unigrid application code plugging into the driver to become AMR-capable.
The driver tasks include the following:
\renewcommand{\labelenumi}{(\roman{enumi})}
\begin{enumerate}
\item For the conserved fluid density $D$, allocate a storage grid function to keep track of the
associated mass correction $\delta M$, {\em i.e.} the total correction to $D$ within the volume 
of a given cell.
\item Upon initialization set all correction arrays $\delta M$ to zero, and compute
the bitmask for the current refinement hierarchy.
\item After any regrid, recompute the bitmask array for the new hierarchy.
\item During the stage when buffer cells are set for variable $D$ at interior boundaries 
on level $L+1$ via interpolation from level $L$, also interpolate 
the correction variable $\delta M$, where the latter's interpolation operator
simply sets $\delta M$ in a child cell to be $1/r^3$ that of the
parent cell (for a three-dimensional spatial grid). 
\item Following injection of arrays $D$ and $\delta M$ from level $L+1$ to level $L$, 
where the injection operator for $\delta M$ is an algebraic sum over child cells (a) zero 
all type B cells in $\delta M$ on level $L+1$, (b) call the application routine 
(first item in the next list) to apply
the mass corrections to $D$ stored in the injected $\delta M$ to type A cells on 
level $L$, (c) zero all type A cells in $\delta M$ on level $L$.
\end{enumerate}
The following are new tasks that the unigrid application code needs to implement:
\begin{enumerate}
\item A routine that will add the mass corrections stored in $\delta M$ to
$D$ for all type A cells on a given grid (i.e.,
set $D_L \rightarrow D_L + \delta M/V_L$)\footnote{Since we consider $D$ 
a density and $\delta M$ a mass, this requires normalization by the volume
element $V_L$, which the application knows. Note that in our code even
though we have included the uncompactified metric volume element $\sqrt{-g}$ 
in the definition of the conservative variables and fluxes, compactification
(and in axisymmetry, the cylindrical shell volume element)
effectively makes the grid non-uniform and so the volume scaling is non-trivial.
An alternative implementation could move this correction step
to the driver list of tasks, though then the application would need 
to supply the driver with the array of local volume elements.}.
\item When taking a single time step on a grid, for any cell marked
type A, set $\delta M$ to minus the
change in mass of the cell from fluxes through cell faces indicated by the bitmask.
For example, with the case illustrated in Fig.\ref{fc_diagram}
and discussed above around Eqs. (\ref{fcL_eqn}) and (\ref{fcL1_eqn}),
set $\delta M_L=-V_L \delta D_{L} $.
\item When taking a single time step on a grid, for any cell marked
type B, add to $\delta M$ the
change in mass of the cell from fluxes through cell faces indicated by the bitmask.
For example, with the same example above,
set $\delta M_{L+1,j,k}\rightarrow \delta M_{L+1,j,k}+V_{L+1,j,k,m} \delta D_{L+1,j,k,m}$.
\end{enumerate}

For the GR-hydro equations we have five conserved fluid variables, $D$ and $S_a$.  
Though the latter do have nonzero source terms --- since gravity can be a source (or sink) of 
energy-momentum --- the above algorithm ensures 
there will be no artificial loss/gain in the presence of AMR boundaries due to 
truncation error from the advection terms.

\section{Tests}
\label{tests}
In this section we present a number of tests of the methods presented above.  
We begin by demonstrating the fourth-order convergence of the
evolution of the Einstein equations for vacuum spacetimes
before moving on to a number of flat space, relativistic hydrodynamics tests
that probe the treatment of fluid discontinuities.  We conclude with several
tests of hydrodynamics in curved spacetimes.  

\subsection{Vacuum evolution}

In~\cite{gh3d,Pretorius:2006tp} several tests of convergence
of an earlier version of the code (without hydrodynamics) were
presented. However, since then we have updated the evolution
of the Einstein equations to fourth-order
spatial differencing and fourth-order Runge-Kutta time differencing,
so we first show two vacuum tests: a Brill wave evolution~\cite{Brill, Sorkin} and a boosted BH evolution.  

\subsubsection{Brill wave}
\begin{figure}
\begin{center}
\includegraphics[width=3.5in,clip=true]{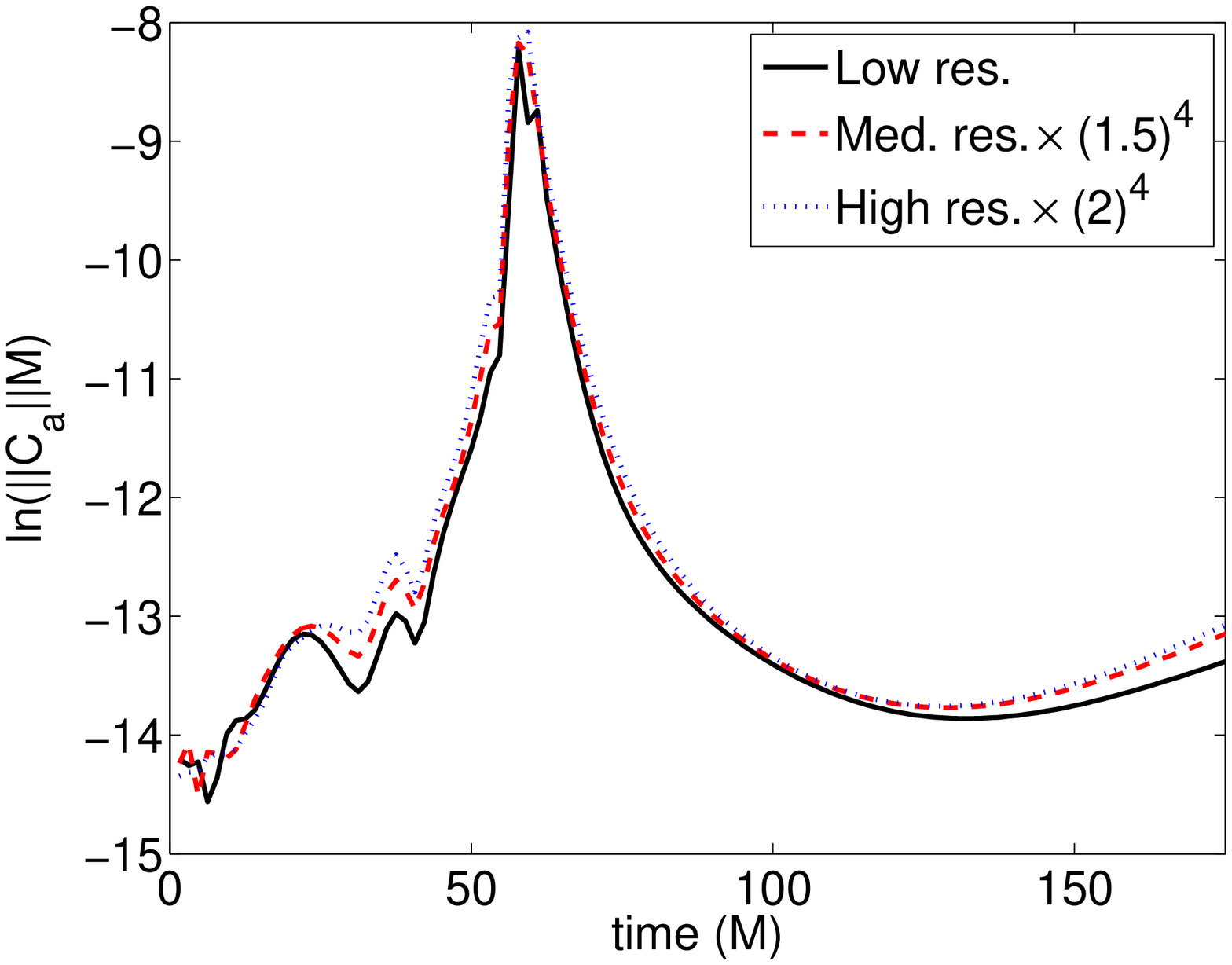}
\includegraphics[width=3.5in,clip=true]{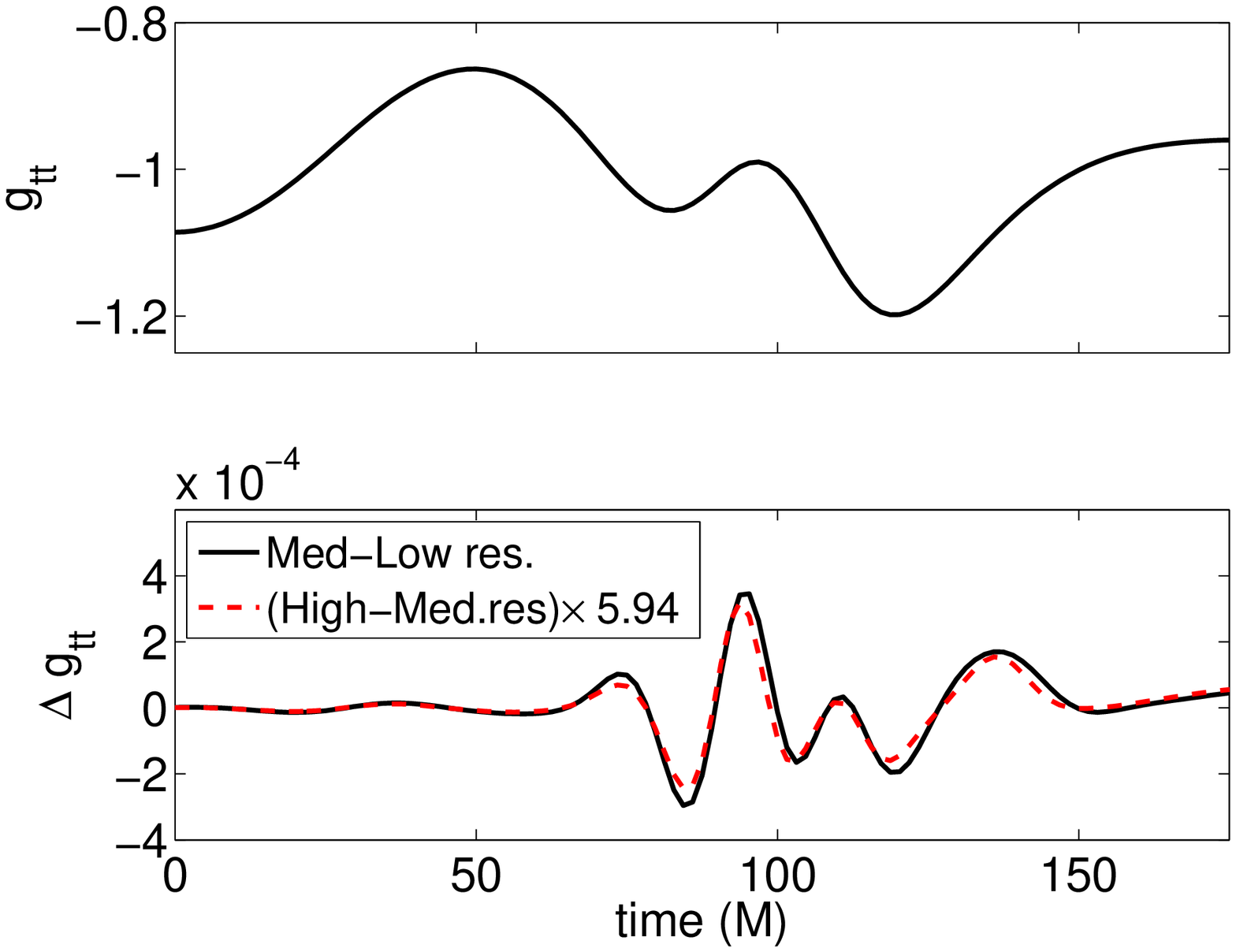}
\caption{{\bf Top}: The natural log of the $L^2$ norm of the constraint violation, $C_a: = H_a-\Box x_a$, for a Brill
wave evolution ({\em i.e.} natural log of $\sqrt{\int|C_a|^2 d^2x / \int d^2x}$).  The three resolutions shown are scaled 
assuming fourth-order convergence.  Time is shown in units of $M$, the total ADM mass of the spacetime, and the constraints are multiplied by $M$ to make them dimensionless. The lowest resolution has a grid spacing of $h=1.56M$.
{\bf Middle/Bottom}: The value of the metric component $g_{tt}$ evaluated at $(x,y,z)=(0,50M,0)$ (middle) and the difference in this quantity between low and medium resolution, and medium and high resolution (bottom), the latter scaled
so that the two curves should coincide for fourth-order convergence. 
\label{bw_const}
}
\end{center}
\end{figure}

For the Brill wave test we begin with initial data
where the spatial line element
is given by 
\begin{eqnarray}
ds^2=\psi^4 \Big(e^Bdx^2+\frac{e^By^2+z^2}{r^2}dy^2+\nonumber\\
\frac{(e^B-1)yz}{r^2}(dydz+dzdy)+\frac{e^Bz^2+y^2}{r^2}dz^2\Big)
\end{eqnarray}
where $r=\sqrt{y^2+z^2}$, $B=2Ar^2\exp(-(r/\sigma_r)^2-(x/\sigma_x)^2)$, and the value of the conformal factor $\Psi$ is determined by solving the Hamiltonian constraint.  We choose $A=40$, $\sigma_r=0.16$, 
and $\sigma_x=0.12$.
The initial data is chosen to be time symmetric ($\dot{\gamma}_{ij}=0$) and maximally sliced
($K = 0$) with the conformal lapse $\tilde{\alpha}:= \Psi^{-6}\alpha=1$.  The remaining metric components are chosen to satisfy the harmonic gauge ($\Box x^a=0$).  This describes a gravitational
wave that initially collapses inward before dispersing.  In Fig.~\ref{bw_const} we show
results from convergence tests in axisymmetry at three resolutions where the medium and high runs had, respectively, 1.5 and 2 times the resolution of the low run.  The constraint equations ($H_a-\Box x_a=0$) as well as the metric components show the expected fourth-order convergence.

\subsubsection{Boosted BH evolution}
\begin{figure}
\begin{center}

\includegraphics[width=3.5in,clip=true]{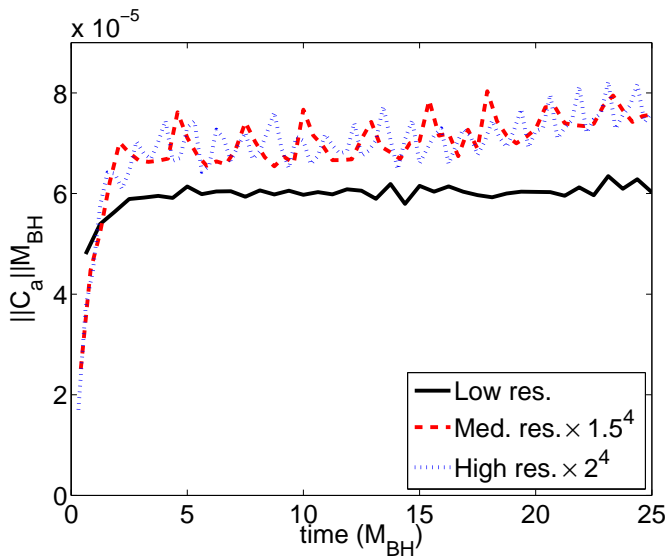}

\caption{
The $L^2$-norm of the constraint violation ($C_a: = H_a-\Box x_a$) in the equatorial plane
for a boosted BH simulation with $v=0.25$.
The three resolutions shown are scaled
assuming fourth order convergence.  Time is shown in units of $M_{\rm BH}$, the ADM mass of the BH in its rest frame,
and the norm of the constraints is multiplied by $M_{\rm BH}$ so as to make it dimensionless.
\label{bbh_cnst}
}
\end{center}
\end{figure}

As an additional vacuum spacetime test we evolved a boosted BH in three dimensions.
We began with initial data describing a BH in harmonic coordinates~\cite{HarmonicBH} with boost parameter $v=0.25$.
As described in~\cite{gh3d}, during the evolution we avoid the BH singularity by searching for an apparent horizon and excising a region within. 
To demonstrate convergence we performed this simulation at three resolutions, the lowest of which has approximately 
30 points covering the diameter of the BH.  The medium (high) resolution has 1.5 (2.0) 
times the number of points in each dimension, respectively.  For all resolutions we used the same AMR hierarchy,
determined based on truncation error estimates at the lowest resolution, with six levels of 2:1 refinement.
In Fig.~\ref{bbh_cnst} we demonstrate that the constraint equations are converging to zero at fourth-order.
When hydrodynamics is included the theoretical limiting convergence rate of our code will
drop to second-order (in the absence of shocks).  However in vacuum dominated regions,
for example the gravitational wave zone, one can expect that for the finite resolutions we
can practically achieve the convergence will be somewhere between second- and fourth-order.

\subsection{Relativistic hydrodynamic tests in flat spacetime}
We have performed a number of standard tests for relativistic,
inviscid hydrodynamics that probe how well a given numerical scheme 
handles the various discontinuities that arise.  The best combination
of reconstruction and flux calculation methods depends on the problem 
under consideration.  We have thus implemented several options and 
maintained a modular code infrastructure so that they are readily 
interchangeable and upgradable.
While strong shocks such as the ones considered here are not expected to play an 
important dynamical role in binary BH-NS mergers, they might be important in other
potential applications of interest (such as NS-NS grazing impacts, or
understanding EM emission from collisions). Thus,
the ability to tailor the reconstruction and flux
methods to the problem at hand may prove important in the future.
In this section, we closely follow the sequence of tests used in the
development of the {\tt RAM} code of Zhang and MacFadyen~\cite{ramcode}, so 
that our results may be compared with theirs.  Though they focus on more
sophisticated flux-reconstruction algorithms, their simpler methods
(labeled U-PPM and U-PLM, denoting reconstruction of the unknowns with
piecewise parabolic and linear reconstruction, respectively) are 
comparable to the ones we employ.  

\subsubsection{1D Riemann problems}  

\begin{table*}
\begin{center}
{\small
\begin{tabular}{ l c c c c c c c }
\hline\hline
 Test & $\Gamma$${}^a$ & $P_L$  &  $\rho_L$ & $v_L$ & $P_R$ &  $\rho_R$ & $v_R$ \\
\hline\hline
  RT1 &  5/3   & 13.33 & 10.0 & 0.0 & $10^{-8}$ & 1.0 & 0.0  \\
  RT2 &  5/3   & 1000.0 & 1.0 & 0.0 & $10^{-2}$ & 1.0 & 0.0  \\
  RT3 &  4/3   & 1.0    & 1.0 & 0.9 & 10.0    & 1.0 & 0.0  \\
  TVT &  5/3   & 1000.0 & 1.0 & 0.0 & 1.0 & 1.0 & $(0.0,0.99)$${}^b$ \\
\hline\hline
\end{tabular}
}
\caption{The initial left and right states for the 1D Riemann problems. \\ 
${}^a$ Adiabatic index of EOS \\ 
${}^b$ In this case $v_x$=0 but the transverse velocity $v_y=0.99$ is nonzero. }
\label{riemann_id}
\end{center}
\end{table*}

\begin{table*}
\begin{center}
{\small
\begin{tabular}{ l l c c c c c c c c }
\hline\hline
Reconstruction & Flux method & \multicolumn{2}{c}{RT1} & 
\multicolumn{2}{c}{RT2} &
\multicolumn{2}{c}{RT3}  & \multicolumn{2}{c}{TVT} \\
\hline
 & & Error${}^a$ & Convergence${}^b$  & Error & Convergence & Error & Convergence & Error & 
Convergence \\
\hline\hline
  MC &  HLL     & 0.034 & 0.82 & 0.110 & 0.59 & 0.062 & 0.77 &  0.238 & 0.72 \\
     & Roe      & 0.032 & 0.82 & 0.110 & 0.60 & 0.052 & 0.80 &  0.233 & 0.72 \\
     & Marquina & 0.036 & 0.82 & 0.127 & 0.59 & 0.056 & 0.79 &  0.227 & 0.76 \\
\hline
  Minmod & HLL  & 0.061 & 0.86 & 0.169 & 0.42 & 0.054 & 0.71  & 0.395 & 0.76 \\
\hline
  WENO5 & HLL   & 0.033 & 0.84 & 0.093 & 0.76 & 0.039 & 0.61 &  0.191 & 0.83 \\
       &  Roe   & 0.032 & 0.85 & 0.096 & 0.79 & 0.039 & 0.60 &  0.198 & 0.81 \\
     & Marquina & 0.036 & 0.85 & 0.093 & 0.76 & 0.038 & 0.66 &  0.183 & 0.82 \\
\hline
  PPM &  HLL & 0.041 & 0.88 & 0.133 & 0.67 & 0.024 & 1.01 & 0.248 & 0.78 \\
\hline\hline
\end{tabular}
}
\caption{1D Riemann test results. \\
${}^a$The L1 norm of the error for resolution $N=400$. \\
${}^b$The average convergence rate between runs with $N=200,$ 400, 800, and 1600.
The ideal rate is unity for problems such as these containing discontinuities.}
\label{riemann_table}
\end{center}
\end{table*}
We first present a series of four relativistic, one-dimensional (1D) Riemann problem tests for which
the exact solution is known (see Sections 4.1-4.4 of~\cite{ramcode}).  In all cases, the domain is
$x\in[0,1]$ and there are initially two fluid states, a left and a right, initially separated
by an imaginary partition at $x=0.5$.  At $t=0$, the partition is removed and the fluid evolves
to some new state.  A $\Gamma$-law EOS is used for all the tests.  In Table~\ref{riemann_id} we 
summarize the initial states and adiabatic indices used for the four tests,
which we label as 
RT1 (Riemann Test 1), RT2, RT3, and TVT (Transverse Velocity Test).  
We compare the performance of the various combinations of reconstruction schemes
and flux methods to the exact solution and summarize the errors and convergence
rates in Table~\ref{riemann_table}.  
Exact solutions to these four
tests were generated using a solver provided by B.\ Giacomazzo, which is 
described in~\cite{bruno_code}.
Taking HLL as our basic flux method, we performed this series of Riemann problem tests
with four reconstruction methods:  MC, minmod, WENO5, and PPM. 
For MC and WENO5, we 
also explored the effect of the flux method by running the tests with the Roe solver and 
Marquina's method.
Most cases have a Courant-Friedrichs-Lewy (CFL) factor of 0.5.
However, the Roe solver, when combined with WENO5, does not seem to work well 
for problems with very strong shocks, such as RT2 and TVT.  For a CFL factor of
0.5, we obtain acceptable results with Roe only by using a more 
diffusive limiter (like MC).  For RT2 and TVT, we thus use
Roe combined with WENO5 with a CFL factor of 0.1.  

\begin{figure}
\begin{center}
\includegraphics[width=3.5in,clip=true]{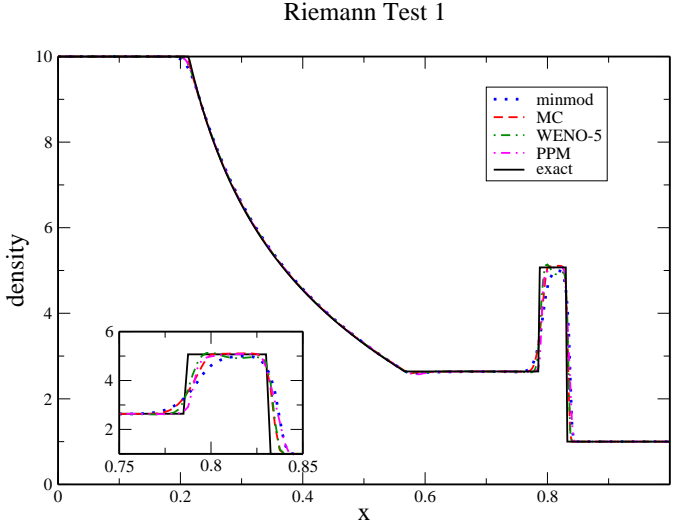}
\caption{Density at $t=0.4$ for Riemann Test 1 (RT1) with
different reconstruction methods and the HLL flux scheme at 
resolution $N=400$.  The inset shows the shock and contact discontinuity.  
The exact solution was generated using the code of~\cite{bruno}. 
\label{rt1_plot}
}
\end{center}
\end{figure}

All of the methods we considered perform well on RT1, which is a fairly easy test.  The lowest overall
error occurs for WENO5 reconstruction (though the density
profile between the shock and the contact discontinuity seems
not to be as flat as in the other cases).  The overall success of WENO5 may be due to 
the fact that the shock is relatively mild and there is an
extended rarefaction that benefits from the high-order
reconstruction.  In Fig.~\ref{rt1_plot} we compare the density profile
obtained using HLL and various reconstruction methods to the exact solution.
We note that the tests which used the Roe or Marquina flux calculation with WENO5 
do not have the oscillation visible in the plot around 
$x=0.8$ in the HLL-WENO5 case.

\begin{figure}
\begin{center}
\includegraphics[width=3.5in,clip=true]{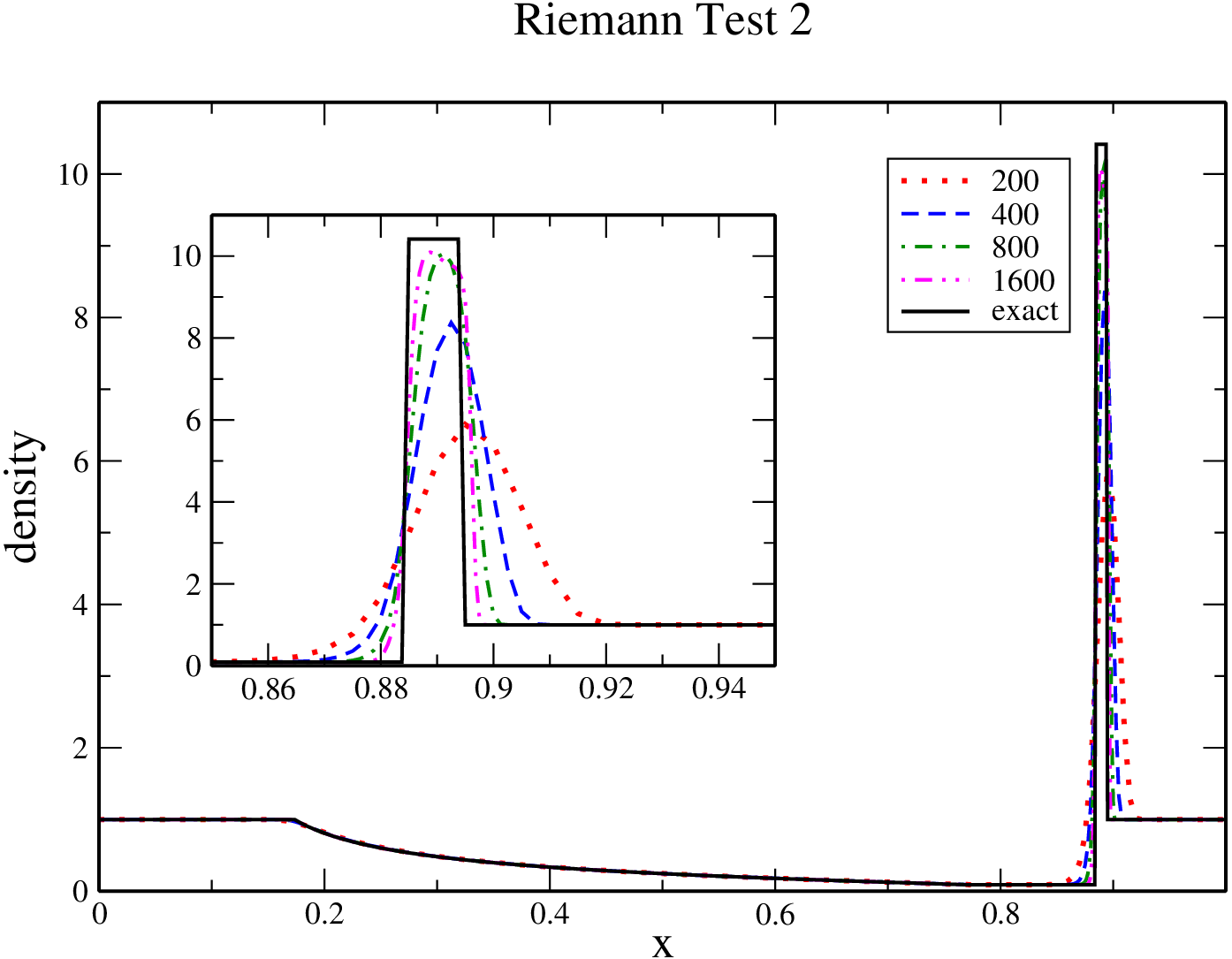}
\caption{Density at $t=0.4$ for Riemann Test 2 (RT2) at different
resolutions for HLL-WENO5.  The average convergence rate in this
case is 0.76.  The thin shell of material between the shock and the
contact is particularly difficult to resolve.
\label{rt2_plot}
}
\end{center}
\end{figure}

The second Riemann test (RT2) is more difficult than the first, with the blast wave 
resulting in a very thin shell of material bounded by a shock 
on the right and a contact discontinuity on the left (see Fig.~\ref{rt2_plot}).  
The average convergence rates for this test show a marked difference 
between the piecewise-linear and higher-order reconstruction 
methods.  WENO5 seems to perform best,
but there is not much difference between HLL and Marquina
or the Roe solver (with diminished CFL factor)
with WENO5.  As in RT1, the reconstruction method seems to be
more important to the solution than the flux scheme.  

\begin{figure}
\begin{center}
\includegraphics[width=3.5in,clip=true]{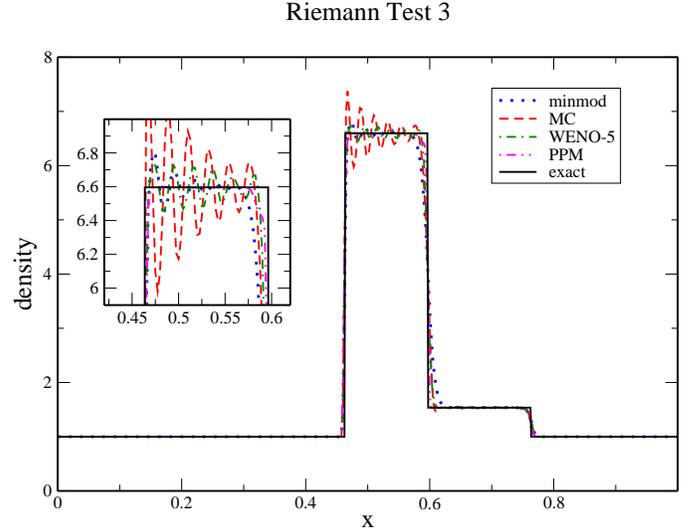}
\caption{Density at $t=0.4$ for Riemann Test 3 (RT3), a collision
problem, for different reconstruction methods with HLL at resolution
$N=400$.  The  post-shock
oscillations are largest in the MC case and smallest for PPM.
\label{rt3_plot}
}
\end{center}
\end{figure}

RT3 is a challenging problem in which the fluid on the left 
collides with the initially stationary fluid on the right, resulting in two
shocks separated by a contact discontinuity.  Our numerical solutions suffer 
from significant oscillations (particularly in the reverse shock) for all 
reconstruction schemes except PPM, which was specifically designed to suppress such 
post-shock oscillations (see Fig.~\ref{rt3_plot}).  PPM also has the
best convergence properties (0.85-1.16), with an average
rate close to the expected value of unity.  (Finite-volume
hydrodynamic schemes such as this should converge at first order
to a weak solution of the equations when discontinuities are
present.) 

\begin{figure}
\begin{center}
\includegraphics[width=3.5in,clip=true]{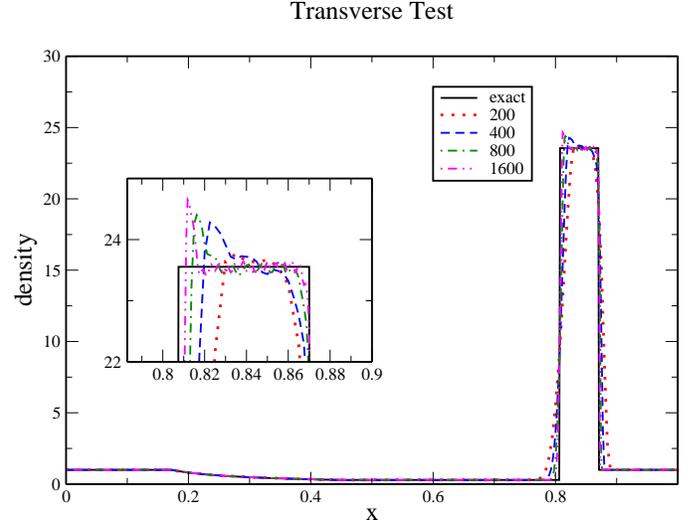}
\caption{Density at $t=0.4$ for the transverse test (TVT) at different
resolutions for HLL-WENO5.  The average convergence rate in this
case is 0.83.
\label{tvt_plot}
}
\end{center}
\end{figure}

For the transverse velocity test, the initial data
are set up as in RT2, except that there is a transverse velocity
$v^y = 0.99$ on the right side of the partition.  
The strong shock propagates {\em into} the boosted fluid,
and the structure of the shock is altered, since the 
velocities in all directions are coupled through the 
Lorentz factor~\cite{PhysRevLett.89.114501}.  Again, the reconstruction technique 
influences the result more than the flux calculation.  
For WENO5 reconstruction, the errors for HLL, Roe, and Marquina are all very close in 
magnitude.  WENO5 and PPM
yield the best results overall.  In Fig.~\ref{tvt_plot} we show 
the density profile at different resolutions for HLL combined with WENO5.  

\subsubsection{1D Shock-heating problem} 

We next consider a one-dimensional shock-heating problem as in~\cite{ramcode}, 
which tests a code's conservation of energy as well as the ability to handle strong 
shocks.  
For this problem, the computational domain is $x\in[0,1]$ with a reflecting
boundary at $x=1$.  The fluid moves toward this boundary with an ultrarelativistic 
initial velocity of $v=1-10^{-10}$.  The fluid has an initial density of $\rho=1.0$ 
and a very small amount of specific internal energy, $\epsilon=0.003$.  The EOS is a gamma-law
with $\Gamma=4/3$.  When the relativistic fluid slams into the wall, its kinetic
energy is converted into internal energy behind a shock which propagates to the left.
Because the fluid is initially cold, essentially all of the heat is generated through
this conversion.  As explained in~\cite{ramcode}, the shock speed and the compression 
ratio of the shock (or equivalently, the post-shock density)
is known analytically.  We evaluate our errors by calculating the L1 norm of the
density errors on the entire computational domain.
The average rate of convergence is also calculated using this measure of error.

\begin{figure}
\begin{center}
\includegraphics[width=3.5in,clip=true]{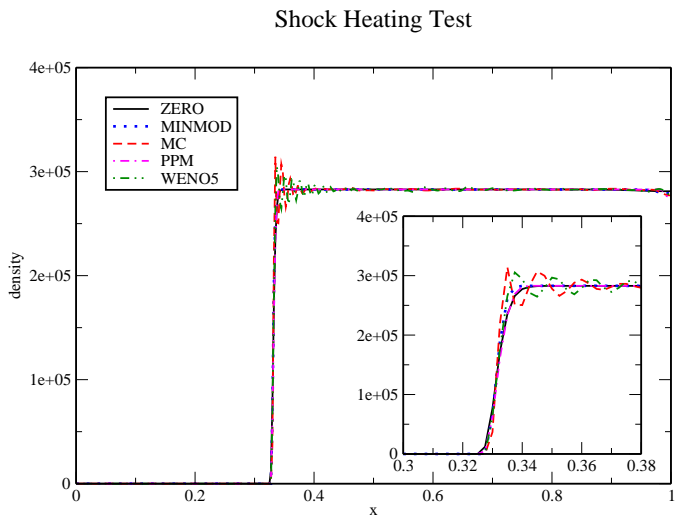}
\caption{Density at $t=2.0$ for the shock-heating test  with
different reconstruction methods and the HLL flux scheme.  The inset 
focuses on the shock front.  
\label{shtplot}}
\end{center}
\end{figure}

We performed this test using HLL with five different reconstruction
methods at four different resolutions (200, 400, 800, and 1600 zones).
Results are shown in Table~\ref{sht_table}.  We find that, due to the
extremely strong shock, there is a tendency for post-shock oscillations
to form with less diffusive reconstruction schemes (see Fig.~\ref{shtplot}).
The WENO5 solution is afflicted with severe post-shock oscillations and exhibits 
poor convergence to the exact compression ratio. Very diffusive reconstruction schemes 
(zero slope and minmod) are comparatively quite successful and converge rapidly
to the exact compression ratio.  PPM, with its flattening step, gives the
best convergence rate overall.  

\begin{table}
\begin{center}
{\small
\begin{tabular}{ l c c }
\hline\hline
Reconstruction & Error${}^a$ & Convergence${}^b$ \\
\hline
  ZERO     & 950  & 0.94 \\ 
  Minmod   & 801  & 0.92 \\ 
  MC       & 1500 & 0.85 \\ 
  PPM      & 824  & 0.96 \\ 
  WENO5    & 1670 & 0.53 \\
\hline\hline
\end{tabular}
}
\caption{Shock -heating test results. For this test we also compare to zero slope
reconstruction, labeled ``ZERO.''  \\
${}^a$ The L1 norm of the error for resolution $N=400$. \\
${}^b$ The average convergence rate between runs with $N=200,$ 400, 800, and 1600. 
}
\label{sht_table}
\end{center}
\end{table}

\subsubsection{Emery step problem}
Next we consider the two-dimensional (2D) Emery step problem~\cite{Emery,ppm}, with the 
setup as in~\cite{ramcode}.
In this scenario, a fluid flows through a wind tunnel at relativistic speed and
hits a step, which is represented by a reflecting boundary condition.  
The computational domain is $(x,y)\in[0,3]\times[0,1]-[0.6,3]\times[0,0.2]$ where the 
subtracted region represents the step.  At the left boundary, inflow conditions are 
enforced (as in the initial data), 
while at the right, outflow conditions are enforced.  All remaining boundaries are 
reflecting. 
The fluid is initialized with density $\rho=1.4$, velocity $v_x=0.999$, and  
a $\Gamma=1.4$ EOS.  The pressure is set to 0.1534, giving a Newtonian Mach number of 3.0.
 
\begin{figure*}
\begin{center}
\includegraphics[width=5in,clip=true]{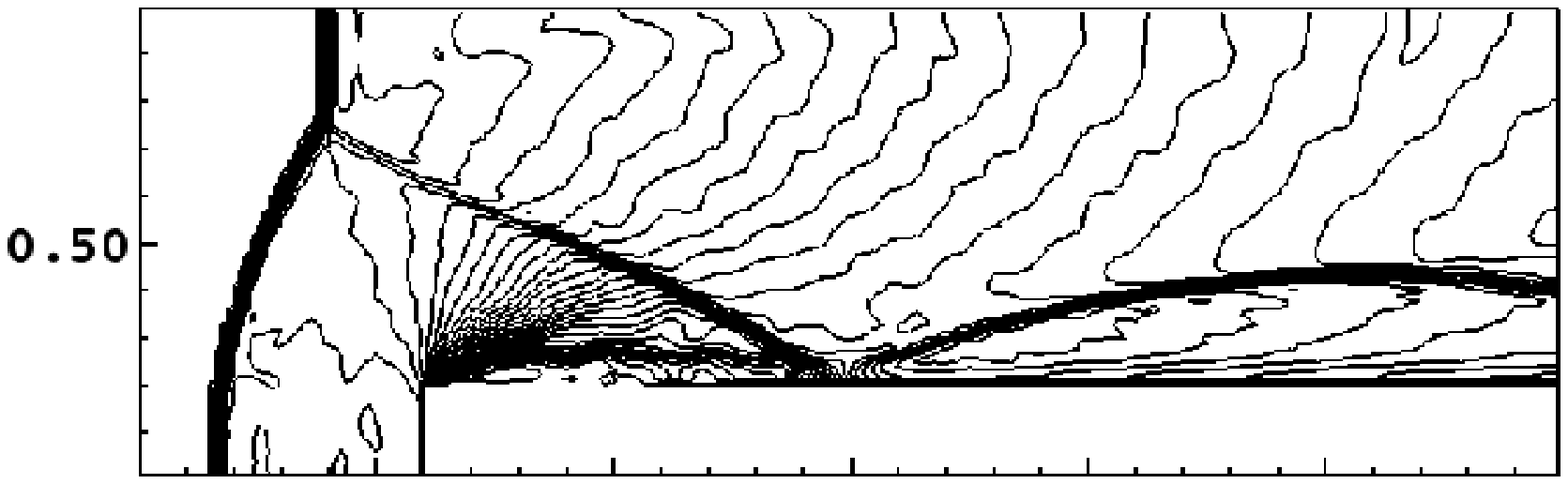} \\
\includegraphics[width=5in,clip=true]{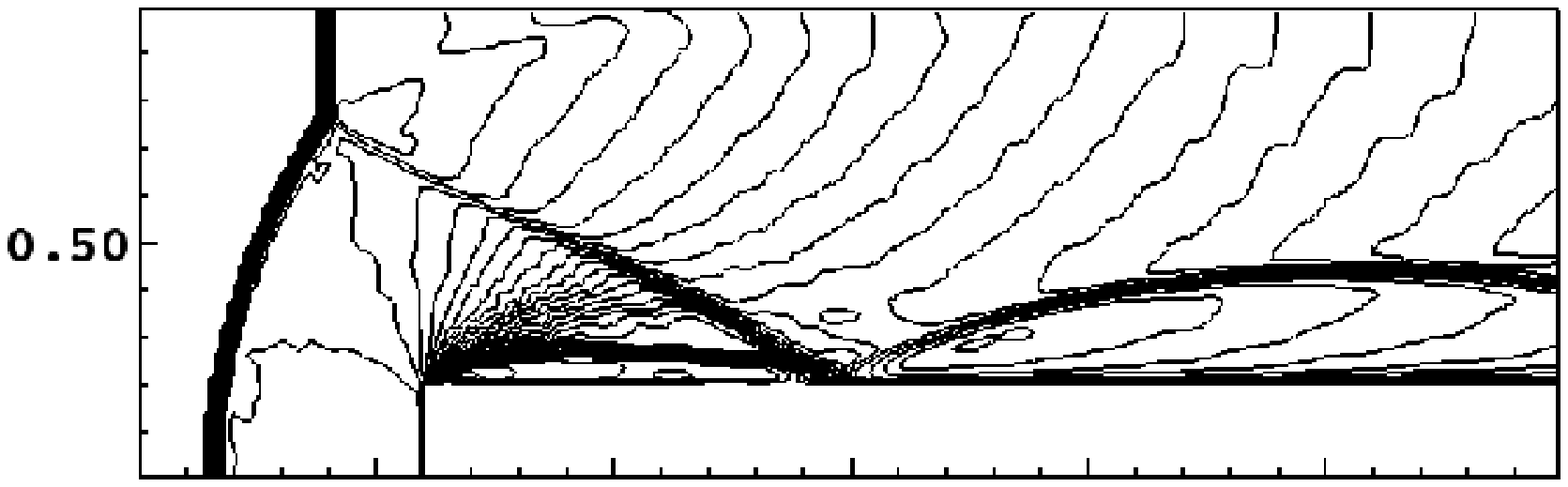} \\
\includegraphics[width=5in,clip=true]{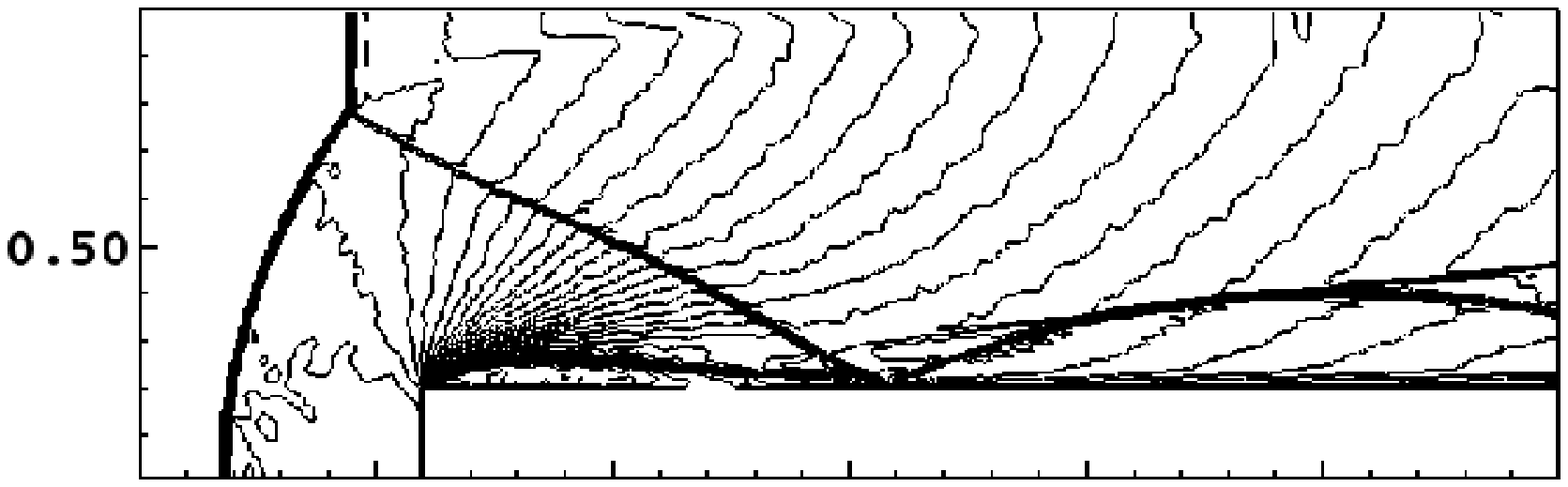} \\
\includegraphics[width=5in,clip=true]{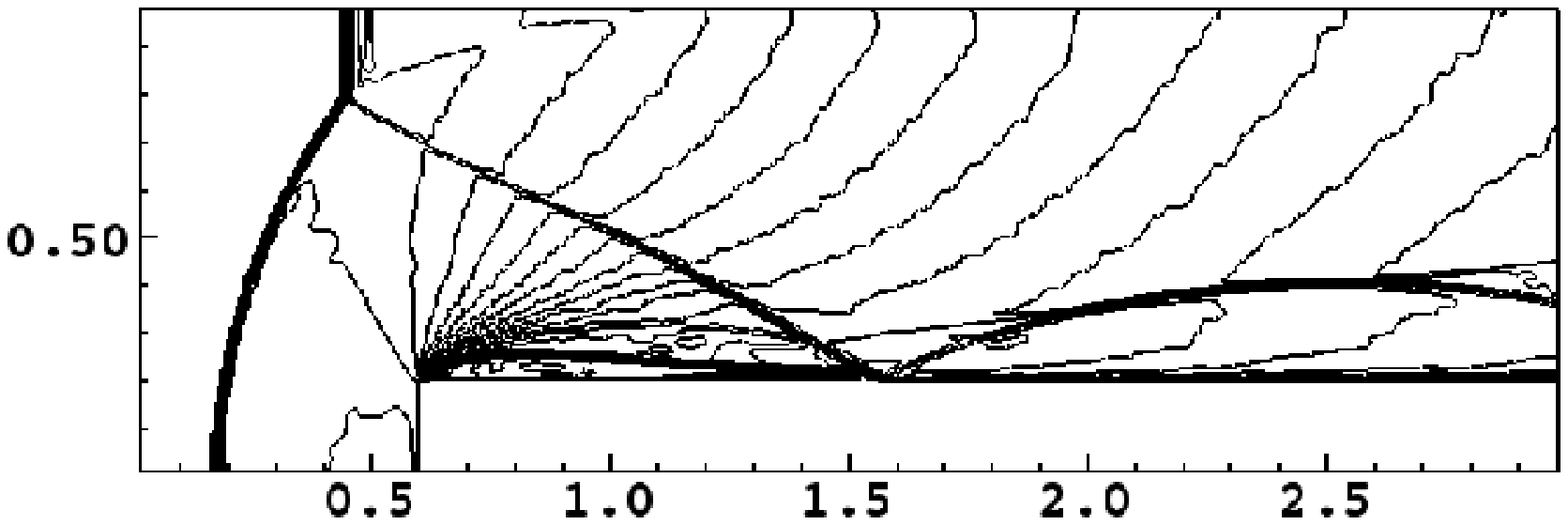}
\caption{Density contours (30 equally spaced in the logarithm)
for the Emery step problem.
The upper (lower) two plots show results for resolution  
$240 \times 80$ ($480 \times 160$).
For each resolution, the upper plot shows results for WENO5
reconstruction, and the lower for PPM.
The respective minimum and maximum densities, $(\rho_{\rm min},\rho_{\rm max})$, are 
$(1.0,1.0\times 10^2)$, $(0.55,1.1 \times 10^2)$, $(0.82,1.1\times 10^2)$, and $(6.8\times 10^{-2},1.1 \times 10^2)$.
\label{emeryfig}
}
\end{center}
\end{figure*}

Higher-order reconstruction methods seem to be essential for this test problem. 
We find that the MC limiter performs poorly, regardless of the flux method.  Although
the MC simulation is stable, the bow shock formed as the fluid reflects off the step
is distorted by large amplitude post-shock oscillations.  These propagate 
downstream, rolling up the boundaries between the different solution regions.
The higher resolution runs with MC also have these features, but at shorter
wavelengths and lower amplitude.  PPM and WENO5 reconstruction performs
much better, and these results are shown at two resolutions in Fig.~\ref{emeryfig}.  
(This figure can be compared to those of~\cite{LucasSerrano:2004aq,ramcode}.)
The PPM results appear slightly better than WENO5 at a given resolution, likely because
of the deliberate oscillation suppression in the PPM algorithm.
 
\subsubsection{2D shock tube problem}

\begin{figure*}
\begin{center}
\includegraphics[width=3.5in,viewport=50 85 570 650, clip=true]{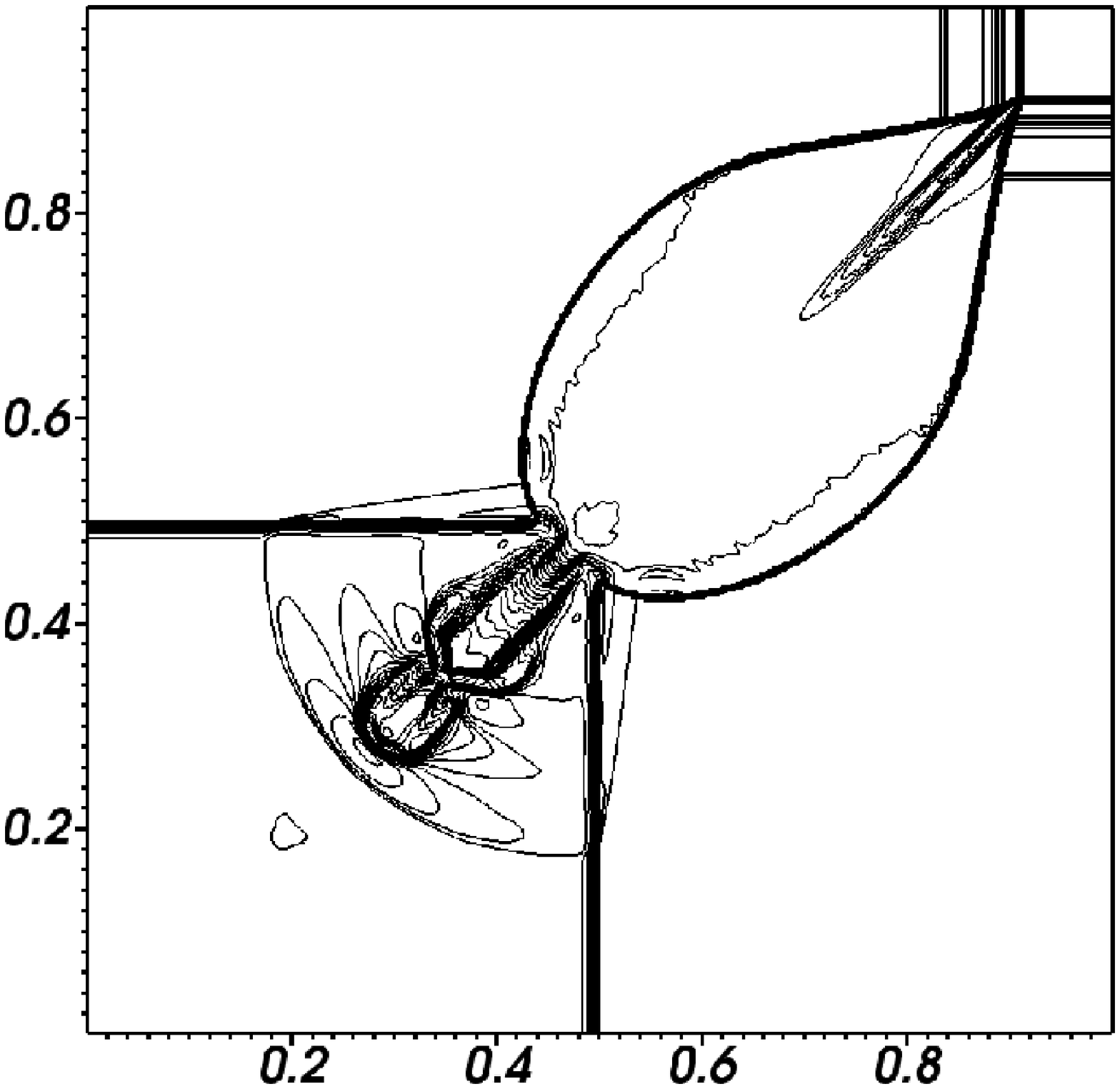}
\includegraphics[width=3.5in,viewport=85 85 605 650, clip=true]{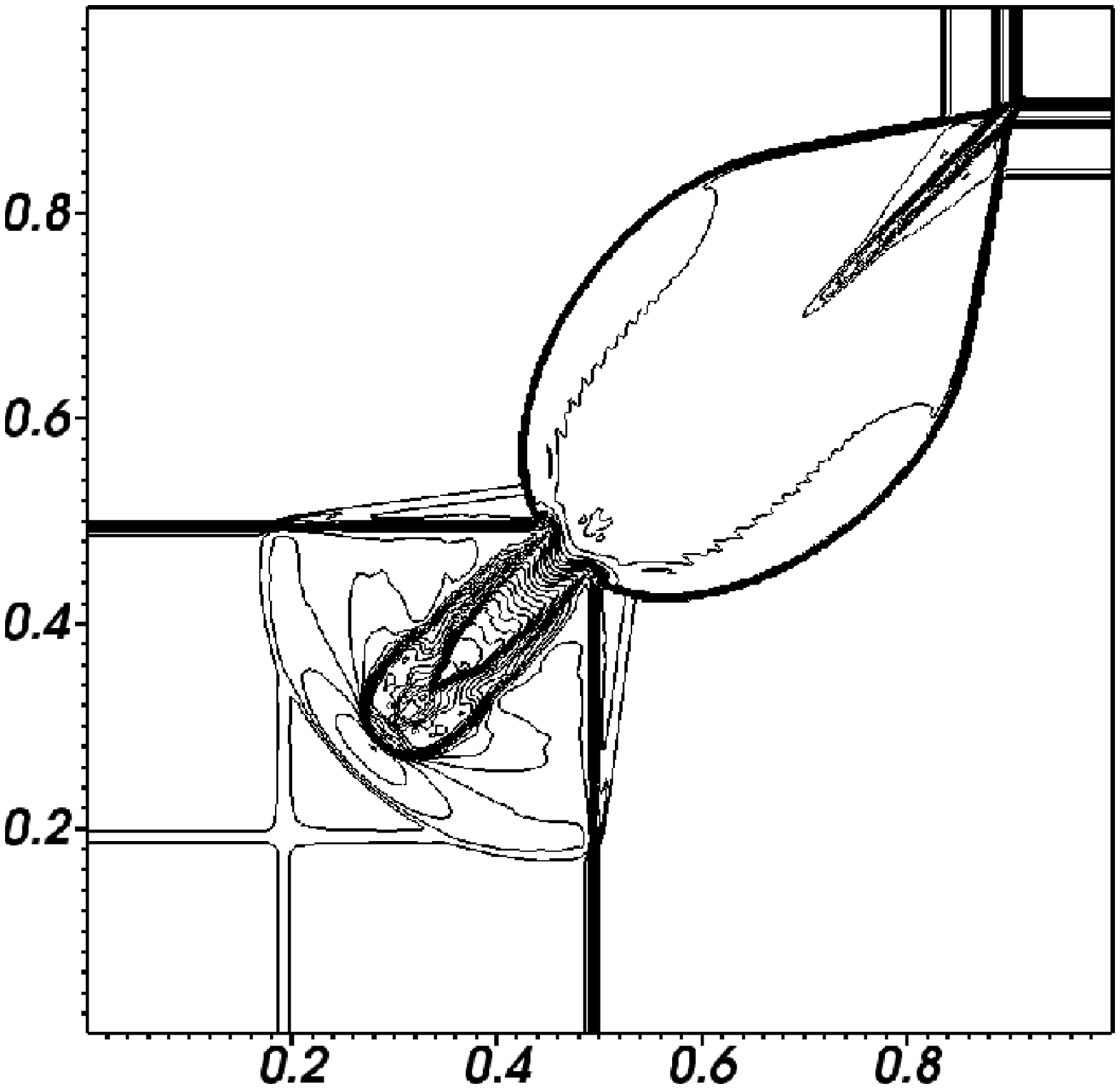} \\
\includegraphics[width=3.5in,viewport=50 50 570 585, clip=true]{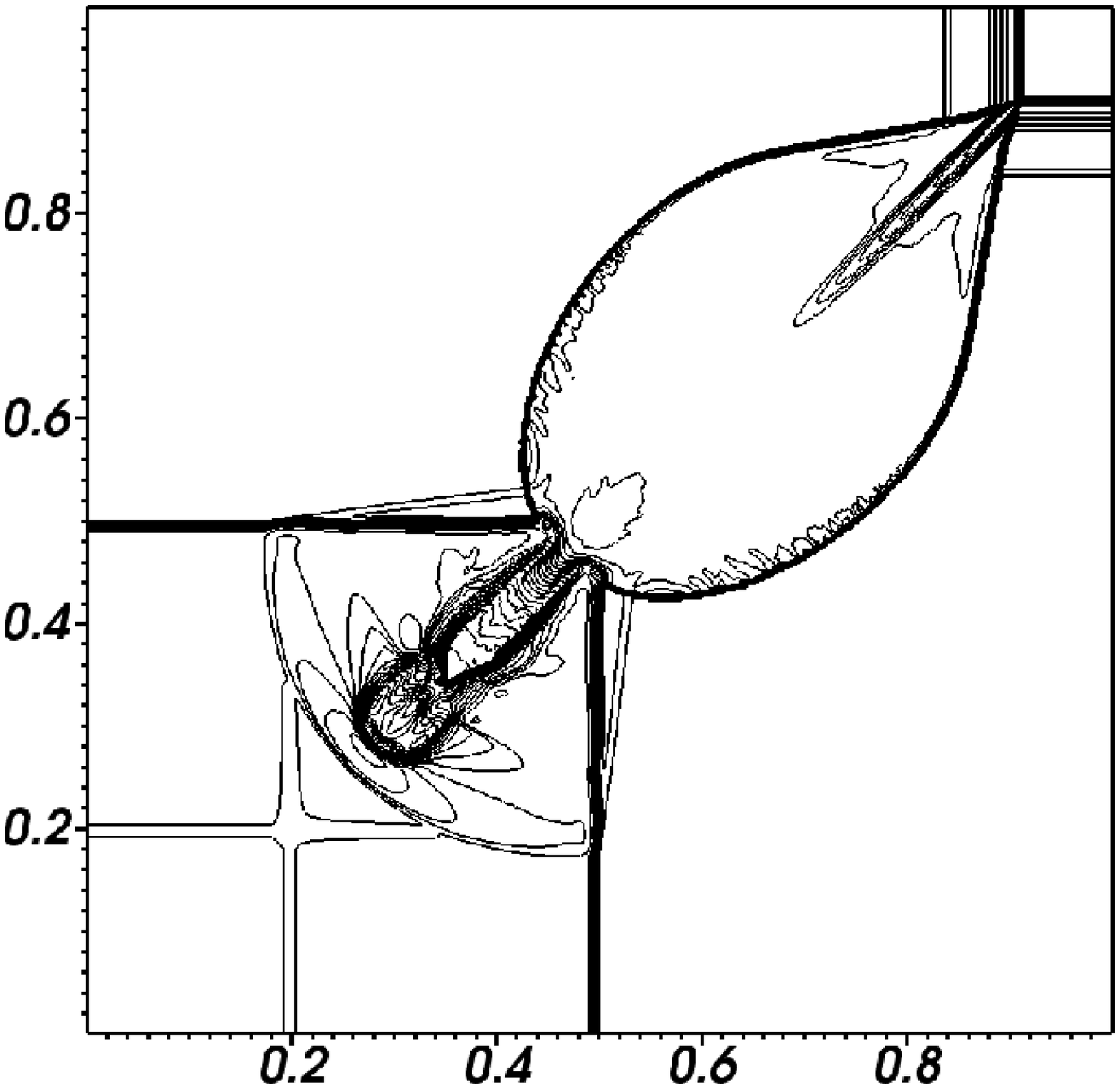} 
\includegraphics[width=3.5in,viewport=85 50 605 585, clip=true]{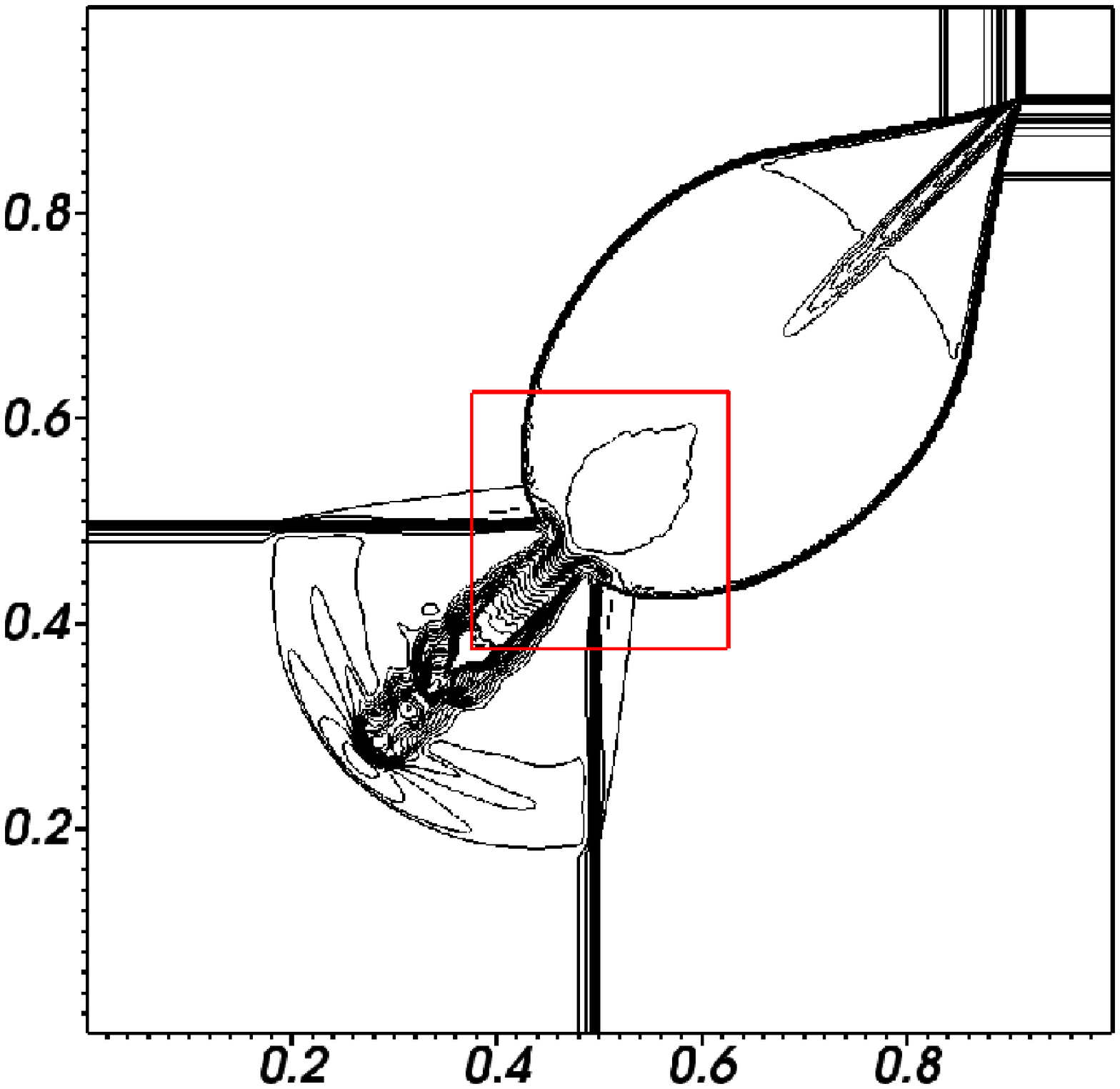}
\caption{Density contours (30 equally spaced in the logarithm)
for the 2D Riemann problem with, from left to right, top to bottom: HLL and MC, HLL and WENO5, Marquina and MC, and HLL and MC with mesh refinement.  
The respective minimum and maximum densities, $(\rho_{\rm min},\rho_{\rm max})$, are 
$(1.1\times 10^{-2},7.0)$, $(8.2\times10^{-3},8.1)$, $(9.1\times 10^{-3},7.1)$, and $(7.6\times 10^{-3},7.0)$.
For the first three simulations a resolution of $400\times400$ was used.  
For the final simulation, a refinement region (red box) was placed in the middle with equivalent resolution, 
while the remaining grid has half the resolution (\emph{i.e.} this simulation has lower resolution overall).  
A CFL factor of 0.5 was used throughout.
\label{r2dfig}
}
\end{center}
\end{figure*}

As an additional test of these algorithms' ability to propagate strong, multidimensional shocks
we consider a 2D shock tube
test.  The computational domain $(x,y)\in[0,1]\times[0,1]$ is divided into four
equal quadrants.  The initial fluid states in the lower/upper, left/right quadrants 
are
\begin{eqnarray*}
(\rho, P, v_x, v_y)^{LL} & = &(0.5,1,0,0)  \\
(\rho, P, v_x, v_y)^{LR} & = &(0.1,1,0,0.99) \\ 
(\rho, P, v_x, v_y)^{UL} & = &(0.1,1,0.99,0)  \\
(\rho, P, v_x, v_y)^{UR} & = &(0.1,0.01,0,0).
\end{eqnarray*}
In this simulation the lower-right and upper-left quadrants converge on the upper-right quadrant creating a pair of curved shocks.
We use a $\Gamma=5/3$ EOS.  In Fig.~\ref{r2dfig} we show results from simulations using HLL or the Marquina flux method combined
with WENO5 or the MC limiter.  The first three panels are from runs with resolution of $400\times400$ and a CFL factor of 0.5 and are comparable
to~\cite{ramcode} and the references therein.  Though the main shock features are captured by all of the methods we considered, 
oscillations arising from the curved shock fronts are evident in varying degrees.  The fourth panel is similar to the first but contains a
refined mesh in 
the center that has the same resolution as the other three panels, while the remainder of the domain has half the resolution.  Though
the majority of the flow is thus effectively derefined, the principal features remain the same.  This is despite the fact that the 
shocks must travel through or along refinement boundaries, and the numerical shock speeds differ slightly on either side of such
boundaries due to the different truncation errors.

\subsection{Hydrodynamic tests in curved spacetime}

\subsubsection{Bondi accretion}
\begin{figure}
\begin{center}
\includegraphics[width=3.5in,clip=true]{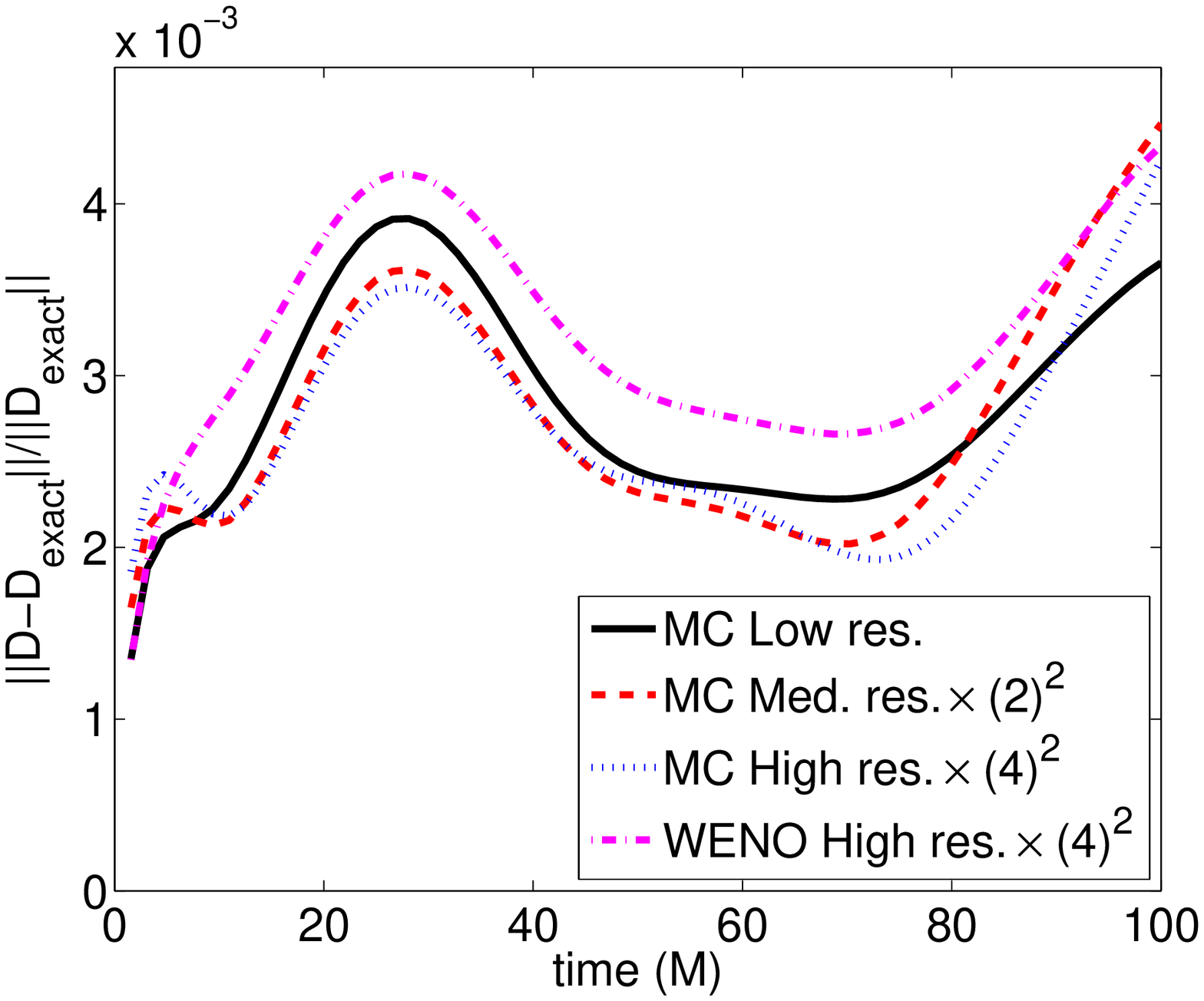}
\caption{
\label{bondi}
The $L^2$ norm of the difference between the numerical and 
exact value of the fluid quantity $D$ (divided by the norm of the exact solution) 
for Bondi accretion with MC and WENO5. The low resolution was run with a grid spacing of $h=0.078 M_{BH}$
while the medium and high has twice and 4 times the resolution 
respectively.  The results are scaled for second-order convergence.
}
\end{center}
\end{figure}

As a first test of our code's ability to simulate relativistic hydrodynamics 
in the strong field regime, we consider Bondi flow.  We set up 
our initial conditions with a stationary solution to spherical accretion onto 
a black hole~\cite{ShapiroTeuk}.  We use Kerr-Schild coordinates for the black
hole metric.  In order to test our code's ability to converge to the correct solution
we measured how the conserved density $D$ differed from the exact solution as a function of time
for three resolutions.  The lowest resolution has a grid spacing of $h=0.078 M_{BH}$, while the medium
and high resolutions have twice and 4 times the resolution respectively.
As shown in Fig.~\ref{bondi}, $||D - D_{\rm exact}||$ converges to zero at second-order.  For this test
we tried both the MC and WENO5 limiters (with HLL for the flux calculation).  
Though both had similar levels of error and showed
the expected convergence, WENO5 had larger errors at low and medium resolutions.  
This is probably because, at lower resolutions, the larger WENO5 stencil extends farther inside 
the black hole horizon where there is larger truncation error. 

\subsubsection{Boosted NS}
\label{bns}
\begin{figure}
\begin{center}

\includegraphics[width=3.5in,clip=true]{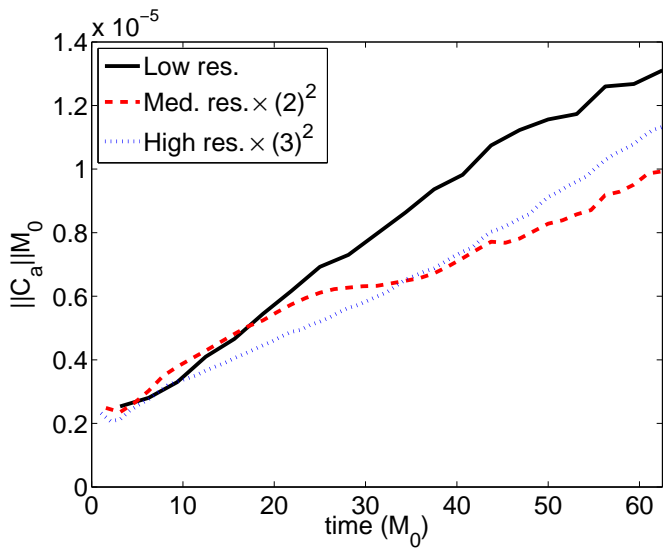}

\caption{
The $L^2$-norm of the constraint violation ($C_a: = H_a-\Box x_a$) in the equatorial plane 
for a boosted NS simulation with $v=0.5$ (using HLL flux calculation and WENO5 limiter).  
The three resolutions shown are scaled 
assuming second-order convergence.  Time is shown in units of $M_0$, the ADM mass of the NS in its rest frame,
and the norm of the constraints is multiplied by $M_0$ so as to make it dimensionless.   
\label{bns_cnst}
}
\end{center}
\end{figure}

\begin{figure}
\begin{center}
\includegraphics[width=3.5in,clip=true]{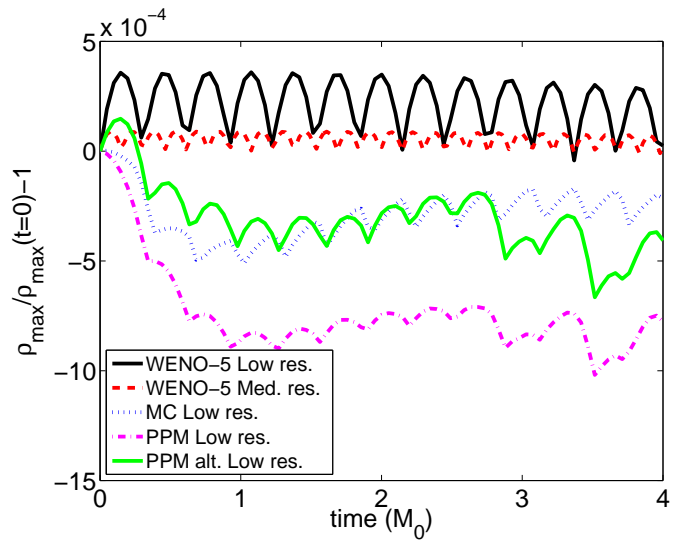}
\caption{
The relative variation of the maximum central density from its initial value 
($\rho_{\rm max}/\rho_{\rm max}(t=0)-1$) during a boosted NS simulation with $v=0.5$
for various reconstruction methods and for two different resolutions in the case of
WENO5.
\label{bns_max_rho}
}
\end{center}
\end{figure}

As an additional test of our evolution algorithm, we considered a single TOV star with a 
boost of $v=0.5$, with astrophysically relevant EOS (the HB EOS of~\cite{jocelyn}) and mass 
(1.35 $M_{\odot}$).
We performed a convergence study at three resolutions, the lowest of which has approximately 50 
points covering the diameter of the star.  The medium (high) resolution has two (three) 
times the number of points in each dimension, respectively. 
The AMR hierarchy is identical in all cases, with 7 levels of 2:1 refinement,
and was determined using truncation error estimates from the low resolution run. 
Figure~\ref{bns_cnst} shows that the 
constraint violations show the expected second-order convergence to zero.  We also compared the 
performance of different reconstruction methods (though using the HLL flux method throughout).  
In Fig.~\ref{bns_max_rho}, we show the maximum density of the NS as a function of time for various 
reconstruction methods.  Though the drifts and oscillations in density converge away for all 
methods, we find that WENO5 gives the least density drift compared to MC and PPM at a given 
resolution.  The drift in maximum density with PPM has to do with the way this particular 
implementation enforces monotonicity at extrema, which results in a loss of accuracy (see 
for example~\cite{new_ppm}).  Modifying the way the algorithm handles smooth extrema can reduce 
this effect.  We implemented one such modification (Eqs. 20-23 from~\cite{new_ppm}), the
results of which are labeled `PPM alt.' in Fig.~\ref{bns_max_rho}.

\subsubsection{Boosted NS flux correction test}
\begin{figure}
\begin{center}
\includegraphics[width=3.5in,clip=true]{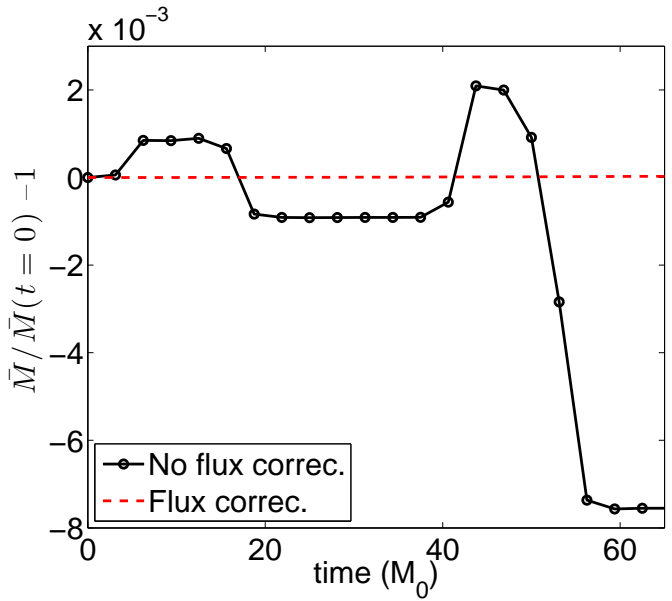}
\caption{
The relative variation of the fluid rest-mass ($\bar{M}/\bar{M}(t=0)-1$) for a boosted NS test
with (dotted, red line) and without (solid black line) flux corrections
at mesh refinement boundaries.  For this test, the mesh hierarchy is fixed 
so that at $t\approx$ 20 $M_0$ the NS has moved from being contained entirely on the finest resolution to being contained entirely on the second
finest resolution.  At $t\approx$ 60 $M_0$ the NS has moved from the second to the third finest resolution.
\label{fc_test}
}
\end{center}
\end{figure}

As a demonstration of the flux correction algorithm (outlined in Sec.~\ref{fc_explain})
to enforce conservation across AMR boundaries, we perform an additional 
boosted NS test.  We use the same conditions as the low resolution simulation outlined above in Sec.~\ref{bns}
but with a different AMR hierarchy.  In particular, we keep the hierarchy fixed 
so that the boosted NS will move to areas of successively lower refinement.  In Fig.~\ref{fc_test},
one sees that without flux corrections there is a $\approx0.1\%$ loss in fluid rest-mass as the NS
moves off the highest refinement level, and a $\approx0.8\%$ loss as the NS moves off the next to highest refinement
level.  This change in the conserved fluid rest-mass comes from the fact that there is a slight mismatch in 
fluxes at the mesh refinement boundaries due to truncation error.  With the flux correction routine
activated, this error is eliminated, and the only change in the total rest-mass is due to the density 
floor criterion ({\em i.e.,} the numerical atmosphere).  
As an indication of how the use of flux corrections affects energy and momentum we
can also compare the integrated matter energy and momentum as seen by a set of Eulerian observers.
The matter energy density is given by $T^{ab}n_an_b$ and the momentum density is given by $p_i=-T^a_in_a$
where $n_a$ is the timelike unit normal to the constant $t$ slices.
These quantities involve combinations of the conserved fluid variables and the metric and are
subject to truncation error, especially since in this simulation the NS is allowed to move
to lower resolution.  In addition, these quantities can vary with time due to gauge effects 
(though in this case, the variation due to gauge effects is subdominant to the variation mentioned below),
so we use them as an indication of the effect of flux corrections by comparing them for
the simulations with and without flux corrections to a similar simulation where the AMR tracks the NS
and there is therefore essentially no flux across AMR boundaries. 
At the end of the simulation the run without flux corrections has $0.91\%$ 
less energy compared to a similar simulation where the AMR tracks the NS, while the run with 
flux corrections has only $0.13\%$ less energy.  The run without flux corrections has $4.0\%$ less momentum 
(in the boost direction) while with flux corrections the comparative loss is $2.5\%$. 

This test is somewhat artificial
since we have deliberately prevented the AMR algorithm from tracking the NS.  As long as the AMR algorithm
can keep the boundaries away from areas of non-negligible flux (as it does when following the boosted NS 
in the test described in Sec.~\ref{bns})
the effect of the flux correction algorithm is small, at the level of the numerical atmosphere that gets pulled along 
with the star. 
However, in astrophysical 
applications, situations may generically arise in which fluid crosses AMR boundaries.  For example, the 
tidal tails formed by the disruption of a NS by a BH will cross refinement boundaries, and
likewise for the subsequent accretion disk that forms, since it would
be much too costly to keep these entire structures on the finest mesh. Of course, the hydrodynamic
solution is still subject to truncation error, which could in principle affect aspects of the dynamics
at the same order of magnitude as putative nonconservative effects. Though for certain
problems, such as calculating the amount of unbound material following a BH-NS merger,
or studying the late time accretion, it could be quite advantageous to ensure conservation
within the hydrodynamic sector. It would be an interesting computational
science problem to systematically study the efficacy of AMR boundary flux conservation in 
such scenarios.

Finally, we note that additional convergence test results from this code were presented 
in~\cite{ebhns_paper} for the particular BH-NS merger simulations discussed there. 

\section{Conclusions}

Numerous scenarios that fall within the purview of general 
relativistic hydrodynamics are still mostly unexplored---especially
CO mergers involving neutron stars.
There is a rich parameter space, of which large areas remain uncharted
due to uncertainty or potential variability in BH and NS masses, BH spin and
alignment, the NS EOS, and other aspects.
Beyond the pure hydrodynamics problem, the roles of magnetic fields and
neutrino physics are just beginning to be explored by various groups,
and we expect to add support for such physics to our code in the future.
The potential applications
of robust and flexible numerical algorithms for evolving hydrodynamics together with 
the Einstein field equations are manifold.  With this in mind, we have implemented methods 
for conservatively evolving
arbitrary EOSs, in particular for converting from conserved to primitive variables without knowledge
of derivatives; and we have implemented numerous reconstruction and flux calculation methods that
can be used interchangeably to meet problem specific requirements.  Though accurate treatment
of shocks may not be crucial for BH-NS mergers (where shocks are not expected to be dynamically
important), the same is not true of NS-NS binaries, especially eccentric ones where
the stars may come into contact during nonmerger close encounters~\cite{Roman}. We have also
taken care to implement a flux correction algorithm that preserves the conservative
nature of hydrodynamical advection across AMR boundaries.  Though strict conservation is not, 
strictly speaking, essential (since any nonconservation would be at the level of truncation 
error), it is an especially appealing property when studying, for example,
CO mergers as potential SGRB progenitors.
After merger, material that did not fall into the black hole --- typically on the order of a
few percent of the original NS mass --- will fill a large volume making up an accretion disk and
potentially unbound material.  Though accurately tracking this material is not important
for the gravitational dynamics, it is critical for characterizing potential EM
counterparts to the merger.

\acknowledgments
We thank Adam Burrows, Matt Choptuik, Luis Lehner, Scott Noble, Inaki Olabarietta,
and Jim Stone for useful conversations.
This research was supported by the NSF
through TeraGrid resources provided by NICS under Grant No.
TG-PHY100053, the Bradley Program (BCS), the NSF
Graduate Research Program under Grant No. DGE-0646086 (WE), NSF Grants No.
PHY-0745779 (FP) and No. PHY-1001515 (BCS),
and the Alfred P. Sloan Foundation (FP).
Simulations were also run on the {\bf Woodhen} cluster
at Princeton University.

\appendix
\section{Spectral decomposition of the flux Jacobian}
\label{spec}
Our conservative formulation of the hydrodynamical equations (\ref{mass_cons},\ref{dtSmu}) can be
written in vector notation as $\partial_t\mathbf{q}+\partial_i(\mathbf{F}^{i}) = \mathbf{S}$ where $\mathbf{q}$ is a five dimensional vector of the conserved (in the absence of sources $\mathbf{S}$) fluid 
variables $\mathbf{q}=(D, S_t, S_x, S_y, S_z)^T$ and the flux
$\mathbf{F}^i=(Dv^i, (S_t-\sqrt{-g}P)v^i, S_jv^i+\delta_j^i\sqrt{-g}P)^T$, where
the index $j$ in the flux is shorthand for the 3 components $(x,y,z)$.
Some flux calculation methods such as the Roe solver~\cite{eulderink} and the Marquina flux~\cite{marquinaFlux} require the spectral decomposition of the Jacobian $\frac{\partial \mathbf{F}^{i}}{\partial \mathbf{q}}$ which we give here. 
(See~\cite{flux_spec} for the spectral decomposition for a similar formulation with slightly different conserved variables.)
The eigenvalues are 
\beq
\lambda_{\pm}=\alpha q(a\pm b)-\beta^i
\eeq
and 
\beq 
\lambda_3=\alpha U^i-\beta^i 
\eeq
(with multiplicity 3), where $a=(1-c_s^2)U^i$,
$b=c_s\sqrt{(1-U^2)[\gamma^{ii}(1-U^2c_s^2)-aU^i]}$,
$q=(1-U^2c_s^2)^{-1}$, $c_s$ is the sound speed,
and $\alpha,\beta^i,\gamma^{ij}$ are metric components
as in (\ref{ADM}).  Here and throughout we use $i\in\{x,y,z\}$ to refer to the direction 
of the flux in the Jacobian with which we are concerned, $\frac{\partial \mathbf{F}^{i}}{\partial \mathbf{q}}$.
In the following equations we use the index $j$ as a shorthand for the three spatial
components of the eigenvectors (that is, the components associated with $S_x$, $S_y$, and $S_z$). The indices $l$ and $m$ are fixed by $i$ and the indices $n$ and $p$ are fixed 
by $j$ as indicated below. The index $k$ is the only index that is summed over.
A set of linearly independent right eigenvectors is given by
\begin{widetext}
\begin{equation}
 \mathbf{r}_{\pm} = \Big(1,hW[U^k\beta_k-\{\alpha(\gamma^{ii}-U^iU^i)+A\beta^i\}/B],hW(U_j-\delta^i_jA/B)\Big)^T,
\end{equation}
\end{widetext}
where
$A=[U^ic_s^2(1-U^2)\mp b]q$ and
$B=\gamma^{ii}-U^i(a\pm b)q$,
\begin{equation}
 \mathbf{r}_{3} = \Big(\kappa/(H W (\kappa/\rho-c_s^2)), U_k\beta^k-\alpha, U_j\Big)^T,
\end{equation}
where $\kappa=\frac{\partial P}{\partial \epsilon}$, and,
\begin{equation}
 \mathbf{r}_{4} = \Big(WU_{l},2hW^2(U_k\beta^k-\alpha)U_l+h\beta_{l}, h(\gamma_{jl}+2W^2U_jU_{l})\Big)^T
\end{equation}
where for $\mathbf{r}_{4}$, $l=y,z,x$ for $i=x,y,z$ respectively.
The expression for $\mathbf{r}_{5}$ can be obtained simply by
replacing $l$ with $m$, where $m=z,x,y$ for $i=x,y,z$ respectively,
in the above expression for $\mathbf{r}_{4}$.
$H$ and $W$ are as defined following (\ref{EnergyEq}).

We also give the  corresponding left eigenvectors.  Component-wise, for $\mathbf{l}_{\pm}=(l^D_{\pm},l^t_{\pm},l^j_{\pm})$,
\begin{widetext}
\begin{eqnarray}
l^D_{\pm}    & = & \mp fhWV_{\mp}\xi \nonumber\\
l^t_{\pm}      & = & \mp f\Big[(K-1)\{-\gamma U^i+V_{\mp}(W^2\xi-\Gamma_{lm})\}+KW^2V_{\mp}\xi\Big]/\alpha \nonumber \\
l^j_{\pm}      & = & \mp f\Big[(\gamma_{ln}\gamma_{mp}-\gamma_{lp}\gamma_{mn})\{1-KA^i-(2K-1)V_{\mp}U^i\}+(2K-1)V_{\mp}\xi W^2U^j\Big]-\beta^jl^t_{\pm}
\end{eqnarray}
\end{widetext}
where $l=y,z,x$ and $m=z,x,y$ for $i=x,y,z$ respectively,
and $n=y,z,x$ and $p=z,x,y$ for $j=x,y,z$ respectively and 
$\Gamma_{lm}=\gamma_{ll}\gamma_{mm}-\gamma_{lm}\gamma_{lm}$,
$\xi=\Gamma_{lm}-\gamma U^iU^i$,
$K=(1-c_s^2\rho/\kappa)^{-1}$, 
$\Lambda_{\pm}=(a\pm b)q$, 
$V_{\pm}=(U^i-\Lambda_{\pm})/(\gamma^{ii}-U^i\Lambda_{\pm})$,
$A^i=(\gamma^{ii}-U^iU^i)/(\gamma^{ii}-U^i\Lambda_{\mp})$, and
\begin{eqnarray*}
f^{-1}&=&2hWbq\xi(K-1)(\gamma^{ii}-U^iU^i)\times\nonumber\\ &&[(\gamma^{ii}-U^i\Lambda_{+})(\gamma^{ii}-U^i\Lambda_{-})]^{-1}.
\end{eqnarray*}

Furthermore,
\begin{equation}
 \mathbf{l}_{3} = \frac{W}{c_s^2\rho}(\kappa-c_s^2\rho)\Big(h, W/\alpha,W(U^j-\beta^j/\alpha)\Big),
\end{equation}
and the components of $\mathbf{l}_{4}$ and $\mathbf{l}_{5}$ are 
\begin{eqnarray}
l^D_{4}    & = & 0 \nonumber\\
l^t_{4}      & = & G_{lm}(\alpha h\xi)^{-1} \nonumber\\
l^j_{4}      & = & \Big[\delta_i^jU^iG_{lm} + \delta_l^j\{\gamma_{mm}(1-U_iU^i)+\gamma_{im}U_mU^i\}\nonumber\\
&&- \delta_m^j\{ \gamma_{lm}(1-U_iU^i)+\gamma_{im}U_lU^i\} \Big](h\xi)^{-1}\nonumber \\ 
-\beta^jl^t_{4} 
\end{eqnarray}
where $G_{lm}= (\gamma_{mm}U_l-\gamma_{lm}U_m)$ and for $\mathbf{l}_{4}$, $l=y,z,x$ and $m=z,x,y$ for $i=x,y,z$ respectively. The
expression for $\mathbf{l}_{5}$ can be obtained from the above expression for $\mathbf{l}_{4}$ simply by interchanging $l$ and $m$.

\bibliographystyle{unsrt}
\bibliography{bhns}

\end{document}